\documentclass[reprint,superscriptaddress, amsmath,amssymb,aps]{revtex4-2}
\usepackage{bm}
\usepackage{colortbl}
\usepackage{graphicx}
\usepackage{hyperref}
\usepackage{siunitx}
\DeclareSIUnit{\molar}{M}

\usepackage[capitalize]{cleveref}
\crefname{section}{Sec.}{Sec.}

\newcommand{\upd}{\mathrm{d}}
\newcommand{\cit}[1]{Ref.~\cite{#1}}
\newcommand{\nn}{\nonumber\\}
\newcommand{\ld}{\lambda}
\newcommand{\kbt}{k_{B}T}

\begin{document}
\title{Direct numerical simulations of the modified Poisson-Nernst-Planck equations for the charging dynamics of cylindrical electrolyte-filled pores}
\date{\today}

\author{Jie Yang}
\affiliation{School of Chemistry and Molecular Engineering, East China University of Science and Technology, Shanghai 200237, China}

\author{Mathijs Janssen}
\email{mathijsj@uio.no}
\affiliation{Department of Mathematics, Mechanics Division, University of Oslo, N-0851 Oslo, Norway}

\author{Cheng Lian}\email{liancheng@ecust.edu.cn}
\affiliation{School of Chemistry and Molecular Engineering, East China University of Science and Technology, Shanghai 200237, China}

\author{Ren\'{e} van Roij}
\affiliation{Institute for Theoretical Physics, Center for Extreme Matter and Emergent Phenomena, Utrecht University, Princetonplein 5, 3584 CC Utrecht, The Netherlands}

\begin{abstract}
Understanding how electrolyte-filled porous electrodes respond to an applied potential is important to many electrochemical technologies. 
Here, we consider a model supercapacitor of two blocking cylindrical pores on either side of a cylindrical electrolyte reservoir. 
A stepwise potential difference $2\Phi$ between the pores drives ionic fluxes in the setup, which we study through the modified Poisson-Nernst-Planck equations, solved with finite elements.
We focus our discussion on the dominant timescales with which the pores charge and how these timescales depend on three dimensionless numbers.
Next to the dimensionless applied potential $\Phi$, we consider the ratio $R/R_b$ of the pore's resistance $R$ to the bulk reservoir resistance $R_b$ and the ratio $r_{p}/\ld$ of the pore radius $r_p$ to the Debye length $\ld$.
We compare our data to theoretical predictions by Aslyamov and Janssen ($\Phi$), Posey and Morozumi ($R/R_b$), and Henrique, Zuk, and Gupta ($r_{p}/\ld$).
Through our numerical approach, we delineate the validity of these theories and the assumptions on which they were based.
\end{abstract}
\maketitle

\section{Introduction}
The dynamics of ions in narrow conducting pores underlies various technologies including biosensors~\cite{TAKHISTOV20041445} and capacitive energy storage~\cite{forse2016jacs,zhan_computational_2017,shao2020}, energy harvesting \cite{PhysRevLett.103.058501}, and water deionization~\cite{D0EE00341G}.
Many of these technologies are based on charging porous electrolyte-filled electrodes, which is a multi-scale process that involves ionic currents over millimetres in electroneutral reservoirs and micron-sized macropores, to form nanometer-sized electric double layers (EDLs) in the electrodes' pores \cite{lian2020blessing}.
Standard electrochemical techniques such as cyclic voltammetry and impedance spectroscopy characterise the response of a macroscopic electrode-electrolyte system \cite{QU199899,eikerling2005optimized,lasia2014electrochemical}. 
The microscopic processes underlying charging of pores, possibly of different size and shape, are then measured all at once; disentangling such microscopic information is not straightforward.
Experimental insight into the charging dynamics at the single-pore level is thus difficult, but progress has been made with nuclear magnetic resonance experiments (albeit on macroscopic porous electrodes)~\cite{wang2017JACS,dou2017nc} and with the surface force balance apparatus \cite{nc2018charging}.
Molecular simulation studies face difficulties opposite to those of experiments as computational power limits simulations to idealised systems of several nanometers at most.
Specifically, many molecular dynamics studies considered ionic liquid-filled slit pores with pore widths comparable to the ion diameters \cite{kondrat2014NM,he2016JPCL,breitsprecher_effect_2017,breitsprecher2018charge,mo2020ion}; cylindrical pores \cite{bi2020molecular} and realistic (but small) porous structures \cite{pean2014dynamics} were also studied. 

These experiments and simulations are often interpreted using the transmission line (TL) model \cite{danielbek1948,ksenzhek1956,levie1963}. 
This model asserts (i) that the charging of a mesoporous electrode filled with dilute electrolyte can be characterised through the charging of a single pore and (ii) that the charging of such a pore can be described by an equivalent circuit, the {\it transmission line circuit}, which distributes the pore's total resistance $R$ and capacitance $C$ over smaller circuit elements. 
In the limit of infinitely many, infinitesimally small resistors and capacitors, the TL circuit gives rise to the differential ``TL equation" [viz.~\cref{eq:TLeq}] for the local electrostatic potential in the pore \cite{janssen2021transmission}.
The TL equation was solved for semi-infinite pores subject to various time-dependent voltages and currents by Ksenzhek and Stender \cite{ksenzhek1956} and de Levie \cite{levie1963}.
They found that a step potential causes the charge $Q$ on the pore to increase with a power law, $Q\propto \sqrt{t}$.
This result can at best represent a short-time regime since, clearly, the charge cannot continue to grow indefinitely. 
Posey and Morozumi \cite{posey1966theory} solved the TL equation for finite-length pores and found that on longer timescales pores charge exponentially with a timescale proportional to $RC$ [see \cref{eq:posey}]. 
These authors also discussed the influence of a bulk reservoir of resistance $R_b$ with which the pore is in contact. 
Gupta and coworkers studied a pore with overlapping EDLs, for which they proposed and solved an amended TL equation [see~\cref{eq:henrique}] \cite{gupta_charging_2020,henrique2021charging}.

Hundreds of articles have used the TL model and its solutions.
Yet, only a handful studied the microscopic physics underlying the TL model---ionic currents in a pore and the EDL formation on its surfaces \cite{sakaguchi2007,lim2009effect,MIRZADEH2014633,PRL2014MTL,gupta_charging_2020,henrique2021charging}.
Sakaguchi and Baba performed direct numerical simulations (DNS) of the Poisson-Nernst-Planck (PNP) equations to study a finite-length pore subject to a suddenly-applied potential \cite{sakaguchi2007}.
These DNS confirmed the short-time power-law scaling but not the exponential relaxation regimes, presumably because ionic charge perturbations did not yet span the entire pore at the latest times they considered [cf. Fig. 1(c) and (d) therein].
DNS of the PNP equations by Mirzadeh, Gibou, and Squires \cite{PRL2014MTL} showed that the TL model accurately describes pore charging for small applied potentials, not only for cylindrical pores but also for other geometries \cite{PRL2014MTL}.
Two recent works further reinforced the TL equation's theoretical basis with first-principles analytical derivations:
both starting from the PNP equations, Henrique and coworkers \cite{henrique2021charging} derived the TL equation and
Aslyamov and Janssen \cite{aslyamov2022analytical} derived the finite-length TL results of Posey and Morozumi.

The TL model only applies to pores subject to applied potentials smaller than the thermal voltage (\SI{24}{\milli\volt} at room temperature).
Several recent articles moved beyond the TL model and studied the response of electrolyte-filled pores subject to larger applied potentials, $\Phi\sim1$, with $\Phi$ the applied potential scaled to the thermal voltage \cite{Robinson_2010,biesheuvel2010nonlinear,PRL2014MTL}.
Robinson, Wu, and Jacobs argued that, at large applied potentials, salt depletion from the pores increases their resistivity, slowing down charging \cite{Robinson_2010}.
Biesheuvel and Bazant also predicted that, after initial TL-model behavior, a slower exponential relaxation sets in with a timescale characteristic of neutral salt diffusion~\cite{biesheuvel2010nonlinear}.
A charging slow-down was indeed visible in the DNS of Mirzadeh and coworkers \cite{PRL2014MTL} with increasing $\Phi$, but the system slowed down less than predicted by \cit{biesheuvel2010nonlinear}.
The authors ascribed this discrepancy to surface conduction: for moderate $\Phi$, the EDLs present a shortcut for ions to bypass the dilute center of the pore.
Semi-analytical results of Aslyamov and Janssen \cite{aslyamov2022analytical} fully agreed with the DNS of \cit{PRL2014MTL}. 

Both the mentioned DNS and analytical derivations concerned the PNP equations, which ignore electrostatic correlations and the finite size of the ions. 
This point-ion approximation is justified for dilute electrolytes and for $\Phi\sim1$, but not for concentrated electrolytes or for larger $\Phi$.
Accordingly, Niya and Andrews studied the charging of porous conductive carbon materials \cite{REZAEINIYA2022139534} through the modified Poisson-Nernst-Planck (MPNP) equations \cite{kilic2007stericII}.
Aslyamov, Sinkov, and Akhatov \cite{aslyamov2022relation} used classical density functional theory to study slit pore charging.
They unified all three known charging regimes:
the pore's charge first increases as if it were semi-infinite ($Q\propto \sqrt{t}$), then slows down and approaches its equilibrium value exponentially with an $RC$ time, and then slows down even further and equilibrates exponentially with the salt diffusion timescale~\cite{aslyamov2022relation}. 

In this article, we report comprehensive DNS of pore charging using the MPNP equations.
We consider many different pore and reservoir sizes, ion diameters, ion concentrations, and applied potentials. 
We focus our discussion on three dimensionless parameters: the ratio $R/R_b$ of the pore's resistance $R$ to the bulk reservoir resistance $R_b$, the ratio $r_{p}/\ld$ of the pore radius $r_p$ to the Debye length $\ld$, and the dimensionless applied potential $\Phi$.
We compare the data from our DNS to theory predictions from Refs.~\cite{posey1966theory,janssen2021transmission,aslyamov2022analytical,henrique2021charging} that have not been tested before.

\begin{figure}
\includegraphics[width=8.6cm]{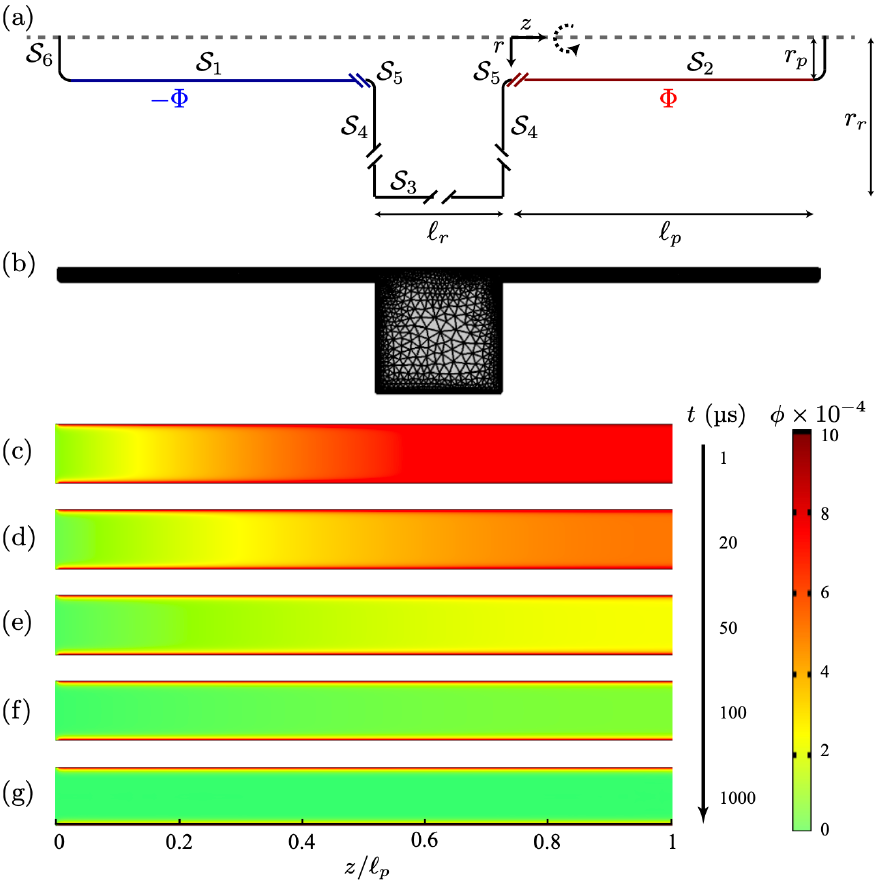}
\caption{
(a) Section view of the microscopic model of two cylindrical pores of length $l_{p}$ and radius $r_{p}$ connected to a cylindrical reservoir of length $\ell_{r}$ and radius $r_{r}$. 
The setup is filled with a 1:1 electrolyte (not shown) with ions of diameter $a$ at salt concentration $c_{b}$.
(b) Representation of a typical mesh to numerically solve the modified Poisson-Nernst-Planck equations.
(c-g) Heat maps of the local electric potential $\phi({\bf r})$ inside the positive electrode pore at times $t$ after switching on a potential $\Phi=10^{-3}$ on the electrode, for (c) $t=\SI{1}{\micro\second}$, (d) $t=\SI{20}{\micro\second}$, (e) $t=\SI{50}{\micro\second}$, (f) $t=\SI{100}{\micro\second}$, and (g) $t=\SI{1000}{\micro\second}$.
We used the ionic diameter $a=\SI{0.1625}{\nano\meter}$, bulk salt concentration $c_b=\SI{0.01}{\molar}$, pore length $\ell_{p}=\SI{1}{\micro\meter}$, pore radius $r_{p}=\SI{50}{\nano\meter}$, and reservoir length and radius $\ell_{r}=r_{r}=\SI{2}{\micro\meter}$.}
\label{fig:setup}
\end{figure}

\section{Model}
\subsection{Setup}
We consider two cylindrical metallic pores of equal length $\ell_{p}$ and radius $r_{p}$ separated concentrically by a cylindrical bulk reservoir of length $\ell_{r}$ and radius $r_{r}$, see \cref{fig:setup}. 
At the ends of the pores are caps of length $r_{p}/5$ with rounded edges of the same radius (the length of the cap is not counted in $\ell_{p}$).
We also add two ``connecting regions" of smooth corners of radius $r_{p}/5$ that link the reservoir to the two pores. 
These regions yield faster convergence of our numerical simulations but have almost no effect on the charging, see \cref{appendix:cap} of the Supplementary Material.
For cases wherein the reservoir and pores have the same radius, we exclude the connecting region between the pore and reservoir.
We denote the surfaces of the two pores by $\mathcal{S}_{1}$ and $\mathcal{S}_{2}$, the boundaries of the reservoir by $\mathcal{S}_{3}$ and $\mathcal{S}_{4}$, and the boundary of the connecting regions and the caps by $\mathcal{S}_{5}$ and $\mathcal{S}_{6}$.
Upon applying a potential between the pores, $\mathcal{S}_{1}$ and $\mathcal{S}_{2}$ will acquire opposite electric charge, while $\mathcal{S}_{3}$ to $\mathcal{S}_{6}$ remain uncharged.
We focus on the charging of the right pore and use a cylindrical coordinate system and a position vector $\mathbf{r}=(r,\theta,z)$ such that $\mathbf{r}=0$ at the left edge of this pore and such that the $z$-axis is aligned with the axes of the pores and reservoir.

The reservoir and pores are filled with a 1:1 electrolyte at a bulk ion concentration $c_{b}$. 
The solvent is treated as a structureless continuum of dielectric constant $\varepsilon=\SI{6.9E-10}{\farad\per\meter}$ and solvent viscosity $\eta=\SI{1.002e-3}{\pascal\per\second}$ (these values are characteristic for water) at a temperature $T=\SI{293}{\kelvin}$.
The cations and anions carry the charge $+e$ and $-e$, with $e$ the elementary charge.
We set ionic diffusivity to $D=\SI{1.34e-9}{\meter\squared\per\second}$, which is typical for alkali halides in water. 
For simplicity, neither the concentration dependence nor the effect of confinement is taken into account for the dielectric constant $\varepsilon$ and the diffusivity $D$. 
For future reference, we define two timescales that will appear repeatedly in our discussion,
\begin{align}\label{eq:deftimescales}
\tau_{I}&= \frac{2\ld}{r_p } \frac{\ell_{p}^2}{D}\,,\quad&\quad
\tau_{II}&= \frac{\ell_{p}^2}{D}\,,
\end{align}
where $\ld=\sqrt{\varepsilon\kbt/(2e^2c_{b})}$ is the Debye length, with $k_{B}$ being Boltzmann's constant. 

As our setup has cylindrical symmetry around the $z$ axis, all physical observables are independent of the azimuthal angle $\theta$. 
We study the time-dependent ionic number densities $\rho_{\pm}(r,z,t)$---the local ionic concentrations scaled to $c_b$---and the dimensionless potential $\phi(r,z,t)$---the local electrostatic potential scaled to the thermal voltage $\kbt/e$.
From $\phi(r,z,t)$, we will determine the right pore's surface charge density 
\begin{equation}\label{eq:surfacechargedensity}
q(z, t)=-\frac{\varepsilon\kbt}{e}\partial_r \phi(\mathbf{r}\in\mathcal{S}_{2}, t)\,,
\end{equation}
and its total surface charge,
\begin{equation}\label{eq:surfacecharge}
Q(t)= 2\pi r_p\int_{\mathcal{S}_{2}} \upd z \, q(z,t) \,.
\end{equation} 
For \cref{eq:surfacechargedensity}, we used that $\mathbf{n}\cdot\bm{\nabla} \phi = \partial_r \phi$ on $\mathcal{S}_{2}$, where $\mathbf{n}$ is the inward normal to the surface.

\subsection{Governing equations}
We model $\rho_{\pm}(r,z,t)$ and $\phi(r,z,t)$ through the MPNP equations,
\begin{subequations}\label{eq:MPNP}
\begin{align} 
\nabla^2 \phi&=-\frac{\rho_{+}-\rho_{-}}{2\ld^2}\,,\label{eq:Poisson} \\
\partial_t \rho_{\pm}&= -\bm{\nabla} \cdot \mathbf{j}_{\pm}\,, \label{eq:continuity}\\
\mathbf{j}_{\pm}&=-D\left[\bm{\nabla} \rho_{\pm}\pm\rho_{\pm} \bm{\nabla} \phi+\frac{a^{3} \rho_{\pm} \bm{\nabla}\left(\rho_{+}+\rho_{-}\right)}{1-a^{3} \left(\rho_{+}+\rho_{-}\right)}\right]\,,\label{eq:mNP}
\end{align}
\end{subequations}
where \cref{eq:Poisson} represents the Poisson equation, \cref{eq:continuity} the continuity equation, and \cref{eq:mNP} the modified Nernst-Planck equation \cite{kilic2007stericII}.
Here, $\mathbf{j}_{\pm}(r, z,t)$ are the ionic fluxes scaled to $c_b$.

We consider the pores to be uncharged and the electrolyte to be homogeneous initially.
At time $t=0$, we apply a positive dimensionless potential $\Phi$ to the right pore and a negative dimensionless potential $-\Phi$ to the left pore. 
This yields the following initial and boundary conditions:
\begin{subequations}\label{eq:MPNPbc}
\begin{align}
\rho_{\pm}(\mathbf{r}, t=0)&=1\,,\\
\phi(\mathbf{r}\in\mathcal{S}_{1},t>0)&=-\Phi\,,\\
\phi(\mathbf{r}\in\mathcal{S}_{2},t>0)&=\Phi\,,\\
\mathbf{n} \cdot\mathbf{j}_{\pm}(\mathbf{r}\in\{\mathcal{S}_{1},\mathcal{S}_{2},\mathcal{S}_{3},\mathcal{S}_{4},\mathcal{S}_{5},\mathcal{S}_{6}\},t)&=0\,,\label{eq:blocking}\\
\mathbf{n} \cdot \bm{\nabla} \phi(\mathbf{r}\in\{\mathcal{S}_{3}, \mathcal{S}_{4},\mathcal{S}_{5},\mathcal{S}_{6}\},t)&=0\,.\label{eq:unchargedwall}
\end{align}
\end{subequations}
Here, \cref{eq:blocking} signifies that all walls are blocking; \cref{eq:unchargedwall} signifies that surfaces of the caps, connecting regions and reservoir boundaries remain uncharged.

\subsection{Numerical implementation}
Numerical simulations for various system parameters $c_{b}, a, \Phi, \ell_{p}, r_{p}, \ell_{r}$, and $r_{r}$ were performed with {\sc comsol multiphysics} 5.4. 
We used a structured nonuniform computational mesh [see \cref{fig:setup}(b)]: coarse in the reservoir domain and finer near all boundaries, where we used a multilayer rectangular grid with a progressively finer layer-to-layer spacing. 
The maximum element size was \SI{10}{\micro\meter}, while the minimum ranged from $0.17$ to \SI{100}{\nano\meter} in the pore domain depending on the Debye length. 
The largest salt concentration we considered was $c_{b}=\SI{0.1}{M}$, for which $\ld=\SI{0.959}{\nano\meter}$.
Hence, the EDL is resolved by at least 5 grid points.

\section{Reservoir-dependent charging}
\subsection{TL model}
As a first example of numerically-determined pore charging, \cref{fig:setup}(c)-(g) shows the dimensionless potential $\phi(r,z,t)$ for five successive times of an electrolyte-filled pore with a bulk concentration $c_{b}=\SI{0.01}{\molar}$ (so that $\ld=\SI{3.03}{\nano\meter}$), ion size $a=\SI{0.1625}{\nano\meter}$, pore length $\ell_{p}=\SI{1}{\micro\meter}$, pore radius $r_{p}=\SI{50}{\nano\meter}$, and reservoir dimensions $\ell_{r}=\SI{2}{\micro\meter}$ and $r_{r}=\SI{2}{\micro\meter}$, subject to a small applied potential $\Phi=10^{-3}$. 
At early times, $\phi(r,z,t)/\Phi=1$ in most of the pore, which implies that the pore's surface charge density and electric field in the pore are both zero.
But near the reservoir, a finite electric field drives counterions into the pore and coions out of it. 
At later times, EDLs form in the nanometer vicinity of the pore surfaces, their width set by the Debye length $\ld$, and the potential $\phi(r,z,t)$ decreases until is zero everywhere except in the EDLs.

The TL model was developed to describe the charging of such pores.
But instead of the full dimensionless potential $\phi(r,z,t)$, the TL equation 
\begin{equation}\label{eq:TLeq}
RC \partial_{t}\psi=\ell_{p}^2 \partial_{z}^{2}\psi\,,\qquad0<z<L\,
\end{equation}
only captures the evolution of $\psi(z,t)=\phi(r=0,z,t)$ at the pore's centerline.
In our case of a cylindrical pore, the pore's resistance amounts to $R = \varrho \ell_{p}/(\pi r_{p}^2)$, with $\varrho=\ld^2/(\varepsilon D)$ the electrolyte resistivity. 
For thin EDLs and small $\Phi$, the pore's Helmholtz capacitance amounts to $C=2\pi r_p \ell_{p} \varepsilon/\ld$.
Their product $RC$ equals $\tau_I$ as defined in \cref{eq:deftimescales}.
For this reason, $\tau_I$ is known as the TL timescale \cite{PRL2014MTL}.
However, this is a bit misleading as the dominant relaxation timescale of a finite-length pore actually also depends on the parameters of the reservoir with which it is in contact \cite{posey1966theory, janssen2021transmission}.
Here, the bulk resistance $R_b$ dependence enters the problem through the boundary conditions to which \cref{eq:TLeq} is subject \cite{posey1966theory, biesheuvel2010nonlinear, janssen2021transmission}, \textit{viz.} 
\begin{subequations}\label{eq:TLeqinitandbcs}
\begin{align}
\psi(z,0)&=\Phi\,,\qquad\qquad0<z<L\,,\label{eq:TLinit}\\
\ell_{p}\partial_{z}\psi(0,t)&=\frac{R}{R_b}\psi(0,t)\,, \label{eq:TLbcRobin}\\
\partial_{z}\psi(\ell,t)&=0\label{eq:TLbcNeumann}\,.
\end{align}
\end{subequations}
Here, \cref{eq:TLinit} describes the initial condition, \cref{eq:TLbcRobin} expresses Kirchhoff's current law at the reservoir-pore interface, and \cref{eq:TLbcNeumann} accounts for the blocking wall at the end of the pore.

For our setup, the bulk resistance $R_{b}=R_r+R_c$ consists of two parts, i.e., the resistance $R_r= \varrho \ell_{r}/(2\pi r_{r}^2)$ of half of the reservoir and the resistance $R_c$ of the connecting region.
This connecting region is bordered by rounded edges of radius $r_p/5$ centered around $z=0$.
The $z$-dependent radius $r_c(z)$ of the connecting region thus satisfies $z^2+(r_c-6r_p/5)^2=r_p^2/5^2$.
To find $R_c$, we view the connecting region as a stack of cylindrical slabs of infinitesimal thickness $dz$ and resistance $\rho dz/A$, with $A=\pi r_{c}^2$.
We then find $R_c=\varrho \int_{-r_p/5}^{0} \upd z /[\pi r_c(z)]^2$, which, upon writing $\bar{z}=z/r_p$, yields
\begin{equation}
R_c=\frac{\varrho}{\pi r_p} \int_{-\frac{1}{5}}^{0} \upd \bar{z} \left(6/5+\sqrt{1/5^2-\bar{z}^2}\right)^{-2}\approx \frac{\varrho}{\pi r_p}\times0.109\,.
\end{equation}
We thus find $R_b=R_r(1+R_c/R_r)$ where $R_c/R_r\approx 0.218 \times r_r^2/(r_p\ell_{r})$.
In our calculations below, this term varies between $R_c/R_r=1.09$ (for $r_r = \ell_{r} = \SI{1}{\micro\meter}$ and $r_p=\SI{200}{\nano\meter}$) and $R_c/R_r=54.5$ (for $r_r = \ell_{r} = \SI{50}{\micro\meter}$ and $r_p=\SI{200}{\nano\meter}$). 
Hence, for very wide reservoirs, the tiny connecting region can constitute the major part of the bulk resistance, $R_b\approx R_c$.
However, the pore's resistance is always vastly larger than that of the connecting region, $R_c\ll R$, so in cases where $R_b\approx R_c$, we have $R_b\ll R$.

Posey and Morozumi solved \cref{eq:TLeq,eq:TLeqinitandbcs} (albeit in different notation) and found \cite{posey1966theory}
\begin{subequations}\label{eq:posey}
\begin{align}
\frac{\psi(z,t)}{\Phi}&=\sum_{j\ge1}\frac{4\sin \beta_j \cos \left[\beta_j\left(1-z/\ell_{p}\right)\right]}{2\beta_j +\sin 2\beta_j} \exp{\!\left(-\frac{t}{\tau_j}\right)}\,,
\intertext{with timescales $\tau_{j}=\tau_{I}/\beta_{j}^2$ and $\beta_j$ solutions of}
\beta_j \tan \beta_j &= \frac{R}{R_b}\,.\label{eq:poseytranscendental}
\end{align}
\end{subequations}

As discussed in \cit{janssen2021transmission}, the early-time charging behavior of the TL equation \eqref{eq:TLeq} is not affected by the the Neumann boundary condition \cref{eq:TLbcNeumann}, which, for all practical purposes, can be taken towards $\ell_p\to\infty$.
A solution to the TL equation for these settings was presented in Eq.~(6) of \cit{janssen2021transmission},
\begin{align}\label{eq:deLevie}
&\frac{\psi(z,t)}{\Phi} =1-\mathrm{erfc}\sqrt{\frac{z^2}{\ell_p^2}\frac{RC}{4t}}\nn
&+\exp\left(\frac{R}{R_b} \frac{z}{\ell_p}+\frac{R^2}{R_b^2} \frac{t}{RC}\right) \mathrm{erfc}\left(\sqrt{\frac{z^2}{\ell_p^2}\frac{ RC}{4t}}+\frac{R}{R_b}\sqrt{\frac{t}{RC}}\right)\,.
\end{align}
We find the total surface charge $Q(t)=-\int_0^{t}\upd t' I(t')$ on the pore, with $I(t)=-(\kbt/e)\ell_p\partial_z\psi(0,t)/R$ the ionic current into the pore, as
\begin{align}\label{eq:Qearly}
Q(t)&=\frac{\kbt}{e}C\Phi\left[\sqrt{\frac{4t}{\pi RC}}-\frac{R_b}{R}\right.\nn
&\quad\left.+\frac{R_b}{R}\exp\left(\frac{R^2}{R_b^2} \frac{t}{RC}\right) \mathrm{erfc}\left(\frac{R}{R_b}\sqrt{\frac{t}{RC}}\right)\right]\,.
\end{align}
When the bulk resistance is negligible compared to the resistance of the pore, $R\gg R_b$, \cref{eq:Qearly} reduces to
\begin{equation}\label{eq:squarerootscaling}
Q(t)=\frac{\kbt}{e}C\Phi\sqrt{\frac{4t}{\pi RC}}\,,
\end{equation}
which is the $\propto \sqrt{t}$ behavior discussed before \cite{sakaguchi2007}.
When the reservoir resistance is not small, $R\sim R_b$,
we find the early-time behavior by expanding \cref{eq:Qearly} for $t/(RC)\ll1$,
\begin{equation}\label{eq:Qearlylinear}
Q(t)=\frac{\kbt}{e}\frac{\Phi}{R_b} \left[ t+O(t^{3/2})\right]\,.
\end{equation}

\begin{figure}
\includegraphics[width=8.6cm]{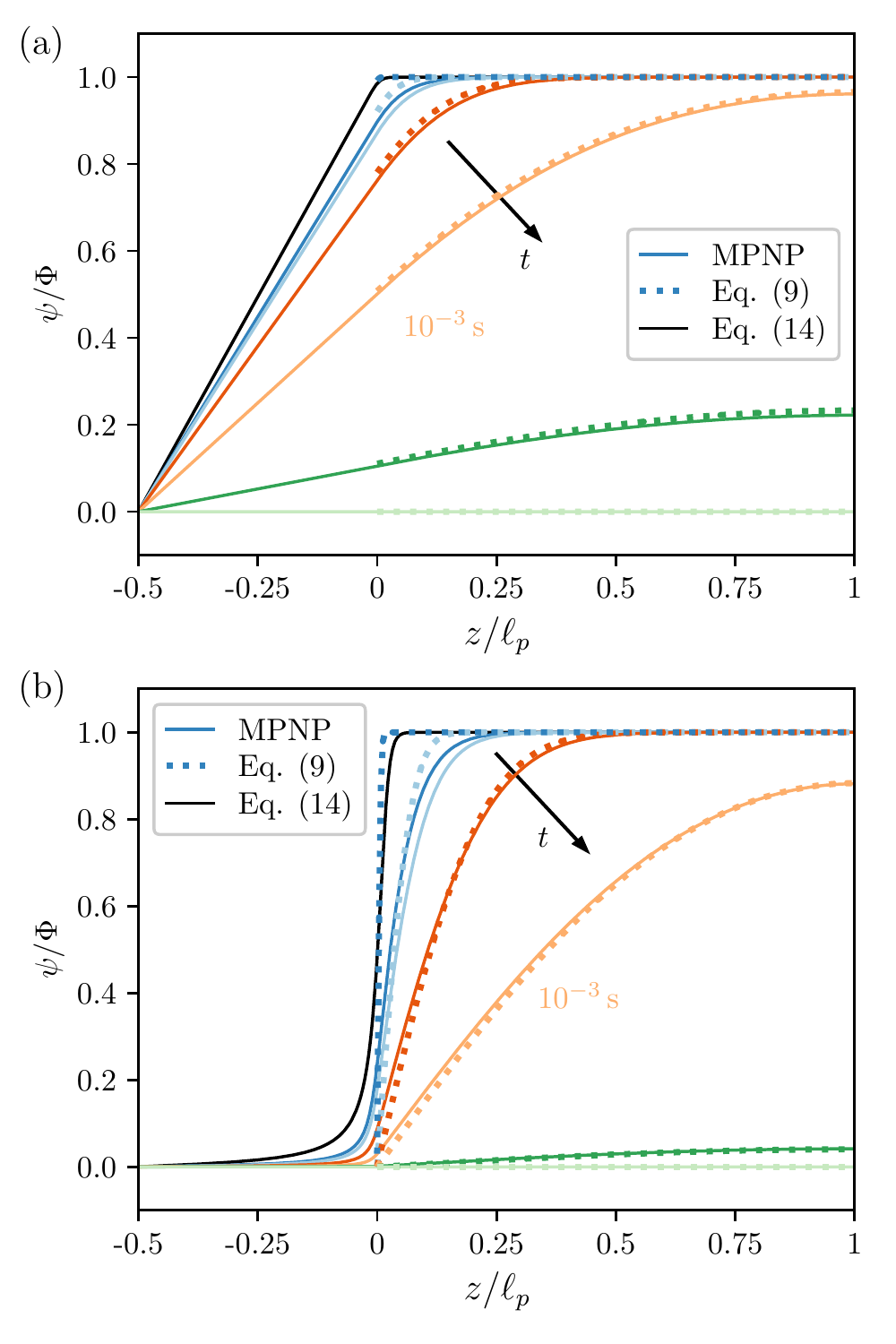}
\caption{
Position dependence of the relative potential on the central axis at different times after switching on a voltage, in (a) for a narrow reservoir $r_{r}=r_{p}$ and in (b) for a wide reservoir $r_{r}=50r_p$ (b). 
The other parameters are set to $\Phi=10^{-3}$, $r_{p}=\SI{200}{\nano\meter}$, $\ell_{p}=l_{r}=\SI{10}{\micro\meter}$, and $c_{b}=\SI{0.001}{\molar}$.
We show numerical solutions to the MPNP equations \eqref{eq:MPNP} for $t=10^{-7}, 10^{-5}, 10^{-4}, 10^{-3}, 10^{-2}, \SI{e-1}{\second}$ (solid lines) and analytical predictions from \cref{eq:posey} (dashed lines) for the same times.
We also show a numerical solution to the Laplace equation \cref{eq:Laplace} (black), which corresponds to $t=0$.}
\label{fig:Laplace}
\end{figure}

\subsection{Comparison of DNS to TL model}
We numerically solve \cref{eq:MPNP,eq:MPNPbc} for a narrow reservoir ($R/R_b=2$) and a wide reservoir ($R/R_b=420.16$) and plot the resulting centerline potential $\phi(r=0,z,t)$ in \cref{fig:Laplace} (solid lines).
In the same figure we plot \cref{eq:posey} (dashed lines).
In both panels we see that, from $t=\SI{e-4}{\second}$ onward, \cref{eq:posey} agrees well with the numerical data although slightly better for the narrower reservoir.
The early times $t=\SI{e-7}{\second}$ and \SI{e-5}{\second} are captured much worse, especially near the pore mouth at $z=0$.
We also show the centerline potential $\phi(r=0,z,t=0)$ (black lines) at the moment of switching on the potential difference.
To determine $\phi(r=0,z,t=0)$, rather than \cref{eq:MPNP,eq:MPNPbc}, we solved the Laplace equation
\begin{subequations}\label{eq:Laplace}
\begin{align} 
\nabla^2 \phi(r, z,t)&=0\,,\\
\phi(\mathbf{r}\in\mathcal{S}_{1},t>0)&=-\Phi\,,\\
\phi(\mathbf{r}\in\mathcal{S}_{2},t>0)&=\Phi\,,\\
\mathbf{n} \cdot \bm{\nabla} \phi(\mathbf{r}\in\{\mathcal{S}_{3}, \mathcal{S}_{4},\mathcal{S}_{5}\},t>0)&=0\,,
\end{align}
\end{subequations}
which is based on the right hand side of the Poisson equation \eqref{eq:Poisson} being zero at $t=0$.
\cref{fig:Laplace}(a) shows that the potential in the reservoir is linear in the special case $r_{r}=r_{p}$, but not if the reservoir is much wider than the pore, as in \cref{fig:Laplace}(b).
Hence, for nontrivial geometries like ours, the Laplace equation is not always solved by a linear potential $\phi(r, z,t)$ in the bulk.
This may have caused the worse performance of the TL model for the wide reservoir, as Refs.~\cite{biesheuvel2010nonlinear,henrique2021charging} motivated \cref{eq:TLbcRobin} by the potential being linear in the reservoir. 
Interestingly, however, \cref{eq:TLbcRobin} can also be derived from the TL circuit \cite{janssen2021transmission}, without any assumption on the potential in the reservoir.
Next, the black line in \cref{fig:Laplace}(b) shows that potential in the pore ($0<z<L$) deviates from $\psi(z=0,t)=\Phi$ at $t=0$.
Hence, the initial condition \cref{eq:TLinit} used in the TL model does not correspond to the numerical simulations.
The discrepancy between \cref{eq:posey} and the MPNP at early times must therefore at least be partially caused by the inaccurate initial condition \cref{eq:TLinit}.

\begin{figure}
\includegraphics[width=8.6cm]{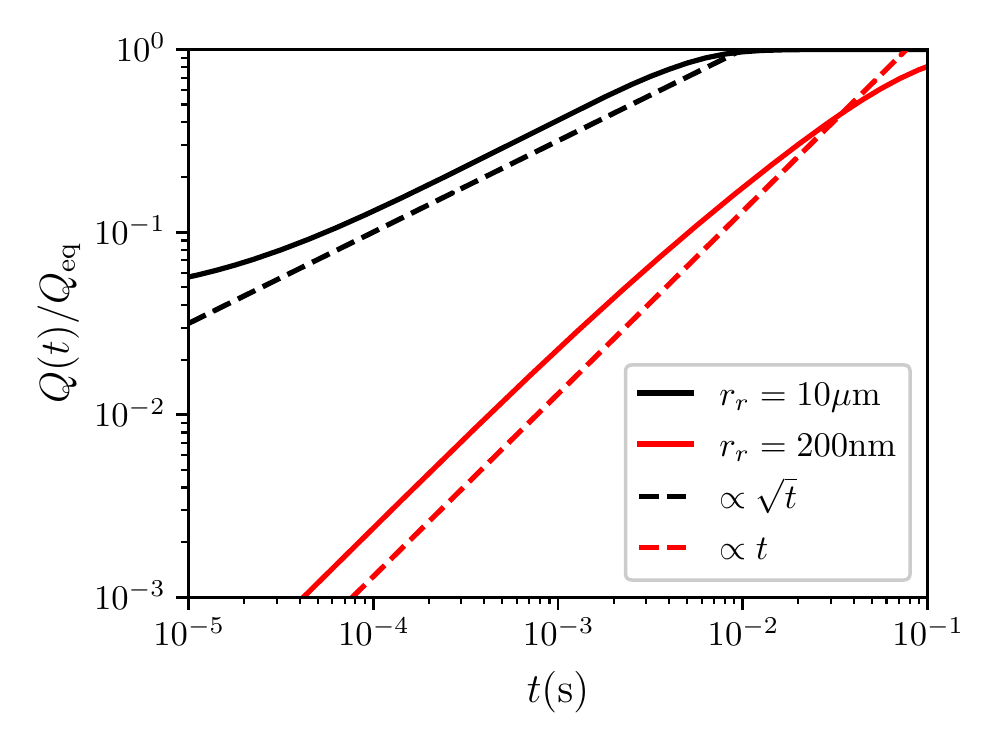}
\caption{
Log-log plots of the normalized surface charge density $Q(t)/Q_{\rm eq}$, for a wide reservoir with $r_{r}=\SI{10}{\micro\meter}$ (black) and a narrow reservoir with $r_{r}=\SI{200}{\nano\meter}$ (red).
The other parameters are set to $c_{b}=\SI{e-3}{molar}$, $\ell_{p}=\SI{10}{\micro\meter}$, $r_{p}=\SI{200}{\nano\meter}$, $\Phi=10^{-3}$, $a=\SI{0.1625}{\nano\meter}$, and $\ell_{r}=\SI{10}{\micro\meter}$.}
\label{fig:law}
\end{figure}

\cref{fig:law} shows the early-time behavior of $Q(t)$ for the same parameters as we used in \cref{fig:Laplace}.
Here, the black line corresponds to $r_{r}=\SI{10}{\micro\meter}$, for which $R/R_b=420.16$, and the red line corresponds to $r_{r}=\SI{200}{\nano\meter}$, for which $R/R_b=2$.
Square-root charging ($Q\propto \sqrt{t}$) is visible for $r_{r}=\SI{10}{\micro\meter}$ up to about $t=\SI{e-2}{\second}$, when the exponential charging starts.
This square-root charging is in line with the theoretical prediction \cref{eq:squarerootscaling} for $R/R_b\gg1$.
For $r_{r}=\SI{200}{\nano\meter}$, the early-time charge accumulation scales linearly, in line with \cref{eq:Qearlylinear} for $R/R_b\sim1$.

\begin{figure}
\includegraphics[width=8.6cm]{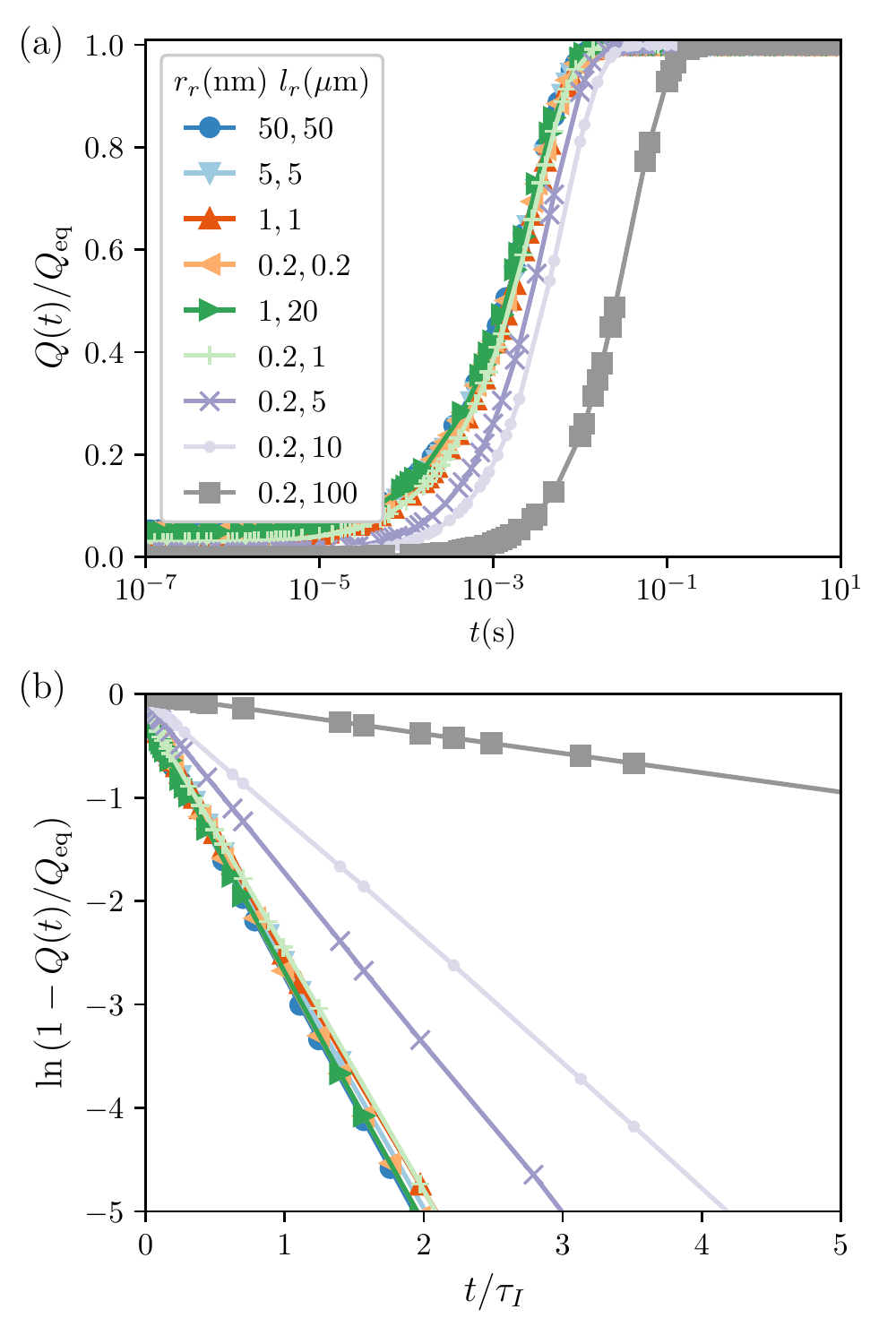}
\caption{
(a) The normalized surface charge density $Q(t)/Q_{\rm eq}$ versus time, and the 
(b) surface charge relaxation versus time scaled by $\tau_{\rm I}$ for different reservoir radii $r_{r}$ and lengths $l_{r}$.
The other parameters are set to $\ell_{p}=\SI{10}{\micro\meter}$, $r_{p}=\SI{200}{\nano\meter}$, $\Phi=10^{-3}$, $a=\SI{0.1625}{\nano\meter}$, $c_{b}=\SI{e-3}{\molar}$.}
\label{fig:res}
\end{figure}

\begin{figure}
\includegraphics[width=8.6cm]{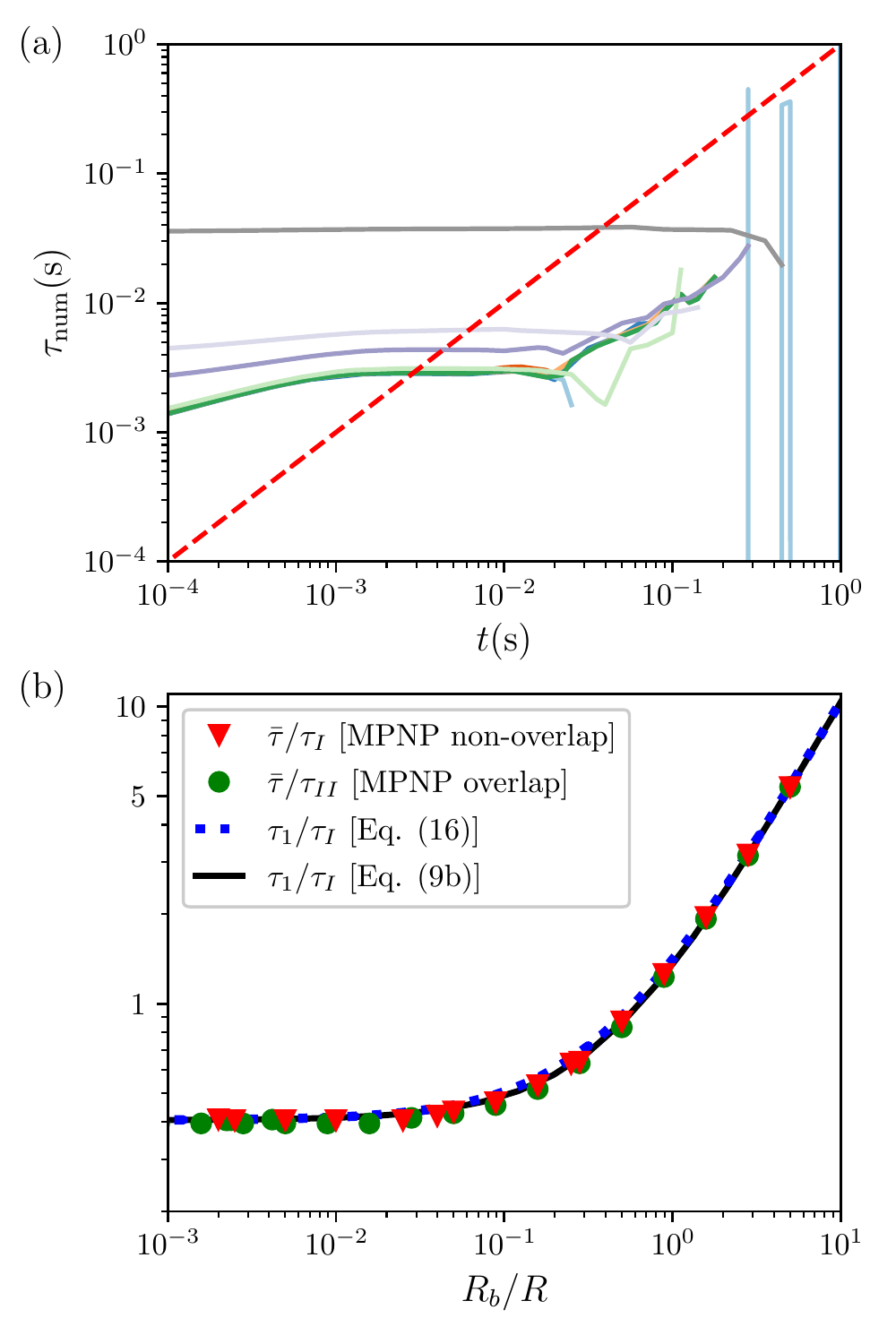}
\caption{
(a) The instantaneous relaxation time $\tau_{\rm num}(t)$ versus time, using the same line styles and parameters as in \cref{fig:res}, and (b) corresponding numerical charging timescale $\bar{\tau}$ scaled by $\tau_{I}$ (red triangles), versus the ratio of reservoir resistance to pore resistance $R_{b}/R$.
The other parameters are set to $\ell_{p}=\SI{10}{\micro\meter}$, $r_{p}=\SI{200}{\nano\meter}$, $\Phi=10^{-3}$, $a=\SI{0.1625}{\nano\meter}$, $c_{b}=\SI{e-3}{\molar}$.
We also show a case with overlapping EDLs (green dots) for which $c_{b}=\SI{e-6}{\molar}$.
Last, we show theoretical predictions from \cref{eq:TLdef} (dashed blue line) and \cref{eq:poseytranscendental}(black line).
We also show a case with overlapping EDLs (green dots) for which $c_{b}=\SI{e-6}{\molar}$.}
\label{fig:res2}
\end{figure}

\subsection{Dependence of the charging time on $R/R_b$}
We further study the dependence of the charging time of pore charging on the size of the reservoir.
\cref{fig:res}(a) shows the normalized surface charge $Q(t)/Q_{\rm eq}$ as a function of time for different reservoir radii $r_{r}$ and lengths $l_{r}$; the legend is arranged in order of increasing $R_{r}=\varrho l_{r}/(2\pi r_{r}^2)$. 
Here, $Q_{\rm eq}$ is the charge $Q(t)$ at the final timestep.
We further set $r_p=\SI{200}{\nano\meter}$ and $c_{b}=\SI{e-3}{\molar}$ such that $\ld/r_p=20.7$; hence, the EDLs are nonoverlapping.
In the figure we see that $Q(t)/Q_{\rm eq}$ does not vanish at $t=0$, which was already suggested by the aforementioned deviations from $\psi(z=0,t=0)=\Phi$ in \cref{fig:Laplace}(b).
Charging relaxation curves overlap for the six smallest $R_r$, implying that the reservoir has no significant influence.
Conversely, for the three largest reservoir resistances, the charging is increasingly slow. 
This slowdown is also visible in \cref{fig:res}(b), where we plot the same data now as $\ln \bm{(}1-Q(t)/Q_{\rm eq}\bm{)}$.
The data in \cref{fig:res}(b) vary linearly versus time on timescales $\tau_{I}$ [\cref{eq:deftimescales}], indicating that the surface charge relaxes exponentially on this timescale.
To characterize this exponential charging in more detail, we introduce the instantaneous numerical relaxation-time function
\begin{equation}\label{eq:Tnum}
{\tau_{\rm num}}(t)=\left[\frac{\upd \ln \bm{(}1-Q(t) / Q_{\rm eq}\bm{)}}{\upd t}\right]^{-1}\,.
\end{equation}
For a purely exponential charging process, $\tau_{\rm num}(t)$ takes a constant value.
In reality, however, $\tau_{\rm num}(t)$ is time dependent:
\cref{fig:res2}(a) shows the instantaneous relaxation time function $\tau_{\rm num}(t)$ \cref{eq:Tnum} for several reservoir radii $r_{r}$ and lengths $l_{r}$ corresponding to the same parameters of \cref{fig:res}.
We see that $\tau_{\rm num}(t)$ grows during the early power-law charging (see \cref{fig:law}) until it reaches a plateau around $t=10^{-3}-\SI{e-1}{\second}$ whose height we denote by $\bar{\tau}$.
(At late times, $Q(t)\approx Q_{\rm eq}$ and the numerical derivative becomes erratic.)
We found that we can effectively determine $\bar{\tau}$ from the intersections of $\tau_{\rm num}(t)$ with $t$ (red dashed) at which time $\tau_{\rm num}=\bar{\tau}$. %
\cref{fig:res2}(b) shows $\bar{\tau}/\tau_I$ vs. $R_b/R$ (red triangles) determined in this way.
We see that $\bar{\tau}/\tau_I$ does not depend on $R_{b}/R$ for small values thereof, and increases linearly with $R_{b}/R$ at large values.
In the same panel, we show the late-time relaxation timescale $\tau_1=\tau_I/\beta_1^2$ of \cref{eq:posey} (black line), for which we numerically solved the transcendental equation \eqref{eq:poseytranscendental}.
Reference~\cite{janssen2021transmission} showed that $\tau_1$ can also be decently approximated by,
\begin{equation}\label{eq:TLdef}
\tau_1\approx RC\left(\frac{4}{\pi^2}+\frac{R_{b}}{R}\right)\,.
\end{equation}
\Cref{fig:res2}(b) shows that both $\tau_1$ determined numerically from \cref{eq:poseytranscendental} and its approximation \cref{eq:TLdef} (blue dashed line) agree well with $\bar{\tau}(t)$.

Instead of $\tau_1$, Posey and Morozumi studied the time at which their $\psi(z=\ell_p,t)$ curve inflected.
Their Fig.~10 of this ``delay time" versus $\log(R_b/R)$ is constant for $R_b/R\ll1$ and increases $R_b/R\gg1$.
Our \cref{fig:res2}(b) [and Fig.~(3) of \cite{janssen2021transmission}] is thus related but not identical to Posey and Morozumi's Fig.~10.

\section{Dependence on EDL overlap $\ld/r_p$}
\subsection{Theory}
Recent work by Fernandez, Zuk, and Gupta \cite{henrique2021charging} generalized the TL model to arbitrary values of $\ld/r_p$.
They found the following centerline potential:
\begin{subequations}\label{eq:henrique}
\begin{align}
\frac{\psi(z,t)}{\Psi}&=I_{0}\left(\frac{r_{p}}{\ld}\right)^{-1}+\left[1-I_{0}\left(\frac{r_{p}}{\ld}\right)^{-1}\right]\nn
&\quad\times\sum_{j\ge1}\frac{4\sin \beta_j \cos \left[\beta_j\left(1-z/\ell_{p}\right)\right]}{2\beta_j +\sin 2\beta_j} 
\exp{\!\left(-\frac{t}{\tau_{j}}\right)}\,,
\intertext{where the timescales $\tau_{j}$ with $j=1,2,\ldots$ read} 
\tau_{j}&=\frac{I_{1}\left(r_{p}/\ld\right)}{I_{0}\left(r_{p}/\ld\right)}\frac{\tau_{I}}{\beta_{j}^2}\,,\label{eq:tauTLhenrique}
\intertext{where $I_0$ and $I_1$ are modified Bessel functions of the first kind, and where $\beta_j$ are the solutions of}
\beta_j \tan \beta_j &= \frac{\ell_p}{\ell_s}\frac{r_s^2}{r_p^2}\,.\label{eq:transcendentalhenrique}
\end{align}
\end{subequations}
In \cref{eq:transcendentalhenrique}, $\ell_s$ and $r_s$ are the length and radius of a ``stagnant diffusion layer" (SDL), a thin region in the reservoir next to the pore over which the potential supposedly drops to zero.
Already noted in \cit{henrique2021charging}, the right hand side of \cref{eq:transcendentalhenrique} is effectively a ratio $R/R_{SDL}$ of the pore resistance to the SDL resistance.
As we did not account for any physical mechanisms (e.g., convection) by which the potential would drop to zero faster than at the center of our reservoir, in the previous section, we preferred using the reservoir size in lieu of the SDL width.
In other words, we prefer replacing $R_{SDL}$ by $R_{b}$.
With this identification, we see that \cref{eq:henrique} reduces to \cref{eq:posey} when $r_{p}\gg\ld$.

\begin{figure}
\centering
\includegraphics[width=8.6cm]{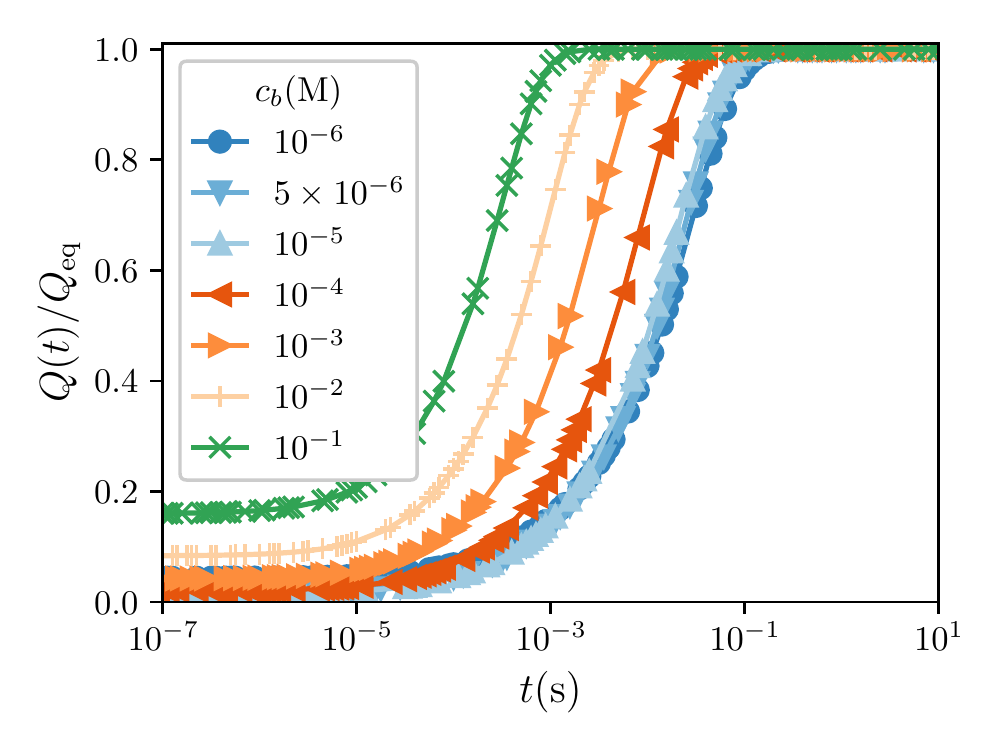}
\caption{The normalized surface charge density $Q(t)/Q_{\rm eq}$ versus time for various electrolyte concentrations of $c_{b}=(10^{-6}-4)\,\si{\molar}$, with $\tau_{\rm I}$ given by \cref{eq:deftimescales}. 
The other parameters are set to $\ell_{p}=\SI{10}{\micro\meter}$, $r_{p}=\SI{200}{\nano\meter}$, $a = \SI{0.1625}{\nano\meter}$, $\Phi=10^{-3}$, $l_{r}=\SI{10}{\micro\meter}$, and $r_{r}=\SI{10}{\micro\meter}$.}
\label{fig:c} 
\end{figure}

Reference \cite{henrique2021charging} already plotted $\psi(z,t)$ from \cref{eq:henrique} vs. $z$ for several times and found good agreement with DNS of the PNP equations. 
Here, we discuss the dependence of the late-time relaxation time $\tau_{1}$ on the various system parameters.

\subsection{$R/R_b$ dependence for $\ld/r_p\gg1$}
When $\ld/r_p\gg1$, we have that $I_{1}\left(r_{p}/\ld\right)/I_{0}\left(r_{p}/\ld\right)\approx r_p/(2\ld)$ so that the late-time relaxation time can be determined from \cref{eq:tauTLhenrique} as $\tau_1=\tau_{II}/\beta_1^2$.
We solved the MPNP equations for different $\ell_r$ and $r_r$ and we set $\ell_p=\SI{10}{\micro\meter}$, $c_b=\SI{e-6}{\molar}$, and $r_p=\SI{200}{\nano\meter}$ so that $r_p/\ld = 0.66$.
From these DNS we determined $\bar{\tau}/\tau_{II}$, which we plot with green dots in \cref{fig:res}.
We see that these scaled data overlap with the $\bar{\tau}/\tau_{I}$ data determined in the previous section for $\ld/r_p\ll1$.

\subsection{$\ld/r_p$ dependence for $R/R_b\gg1$}
Next, we considered many different $c_b, r_p, \ell_p$, and $a$.
In all cases, $R/ R_b\gg1$; the smallest value considered was $R/R_b\approx193.80$.
For such large $R/R_b$, we can use that in the limit of $R/ R_b\to\infty$, \cref{eq:transcendentalhenrique} is solved by $\beta_{1}=\pi/2$ and 
\begin{equation}\label{eq:tau1RRbinf}
\tau_{1}=\frac{4}{\pi^2}\frac{I_{1}\left(r_{p}/\ld\right)}{I_{0}\left(r_{p}/\ld\right)}\tau_I\,.
\end{equation}

First, we investigate how the electrolyte concentration affects the charging dynamics. 
\cref{fig:c} shows the surface charge density $Q(t)/Q_{\rm eq}$ versus time for several $c_b$. 
We see that charging goes faster at higher electrolyte concentration, which agrees with the $\tau_{I}$ timescale [\cref{eq:deftimescales}] from TL theory. 
Moreover, this panel shows that the charge data collapses for concentrations below \SI{e-5}{\molar}, for which Debye lengths are comparable or larger than the pore radii. 

\begin{figure}
\includegraphics[width=8.6cm]{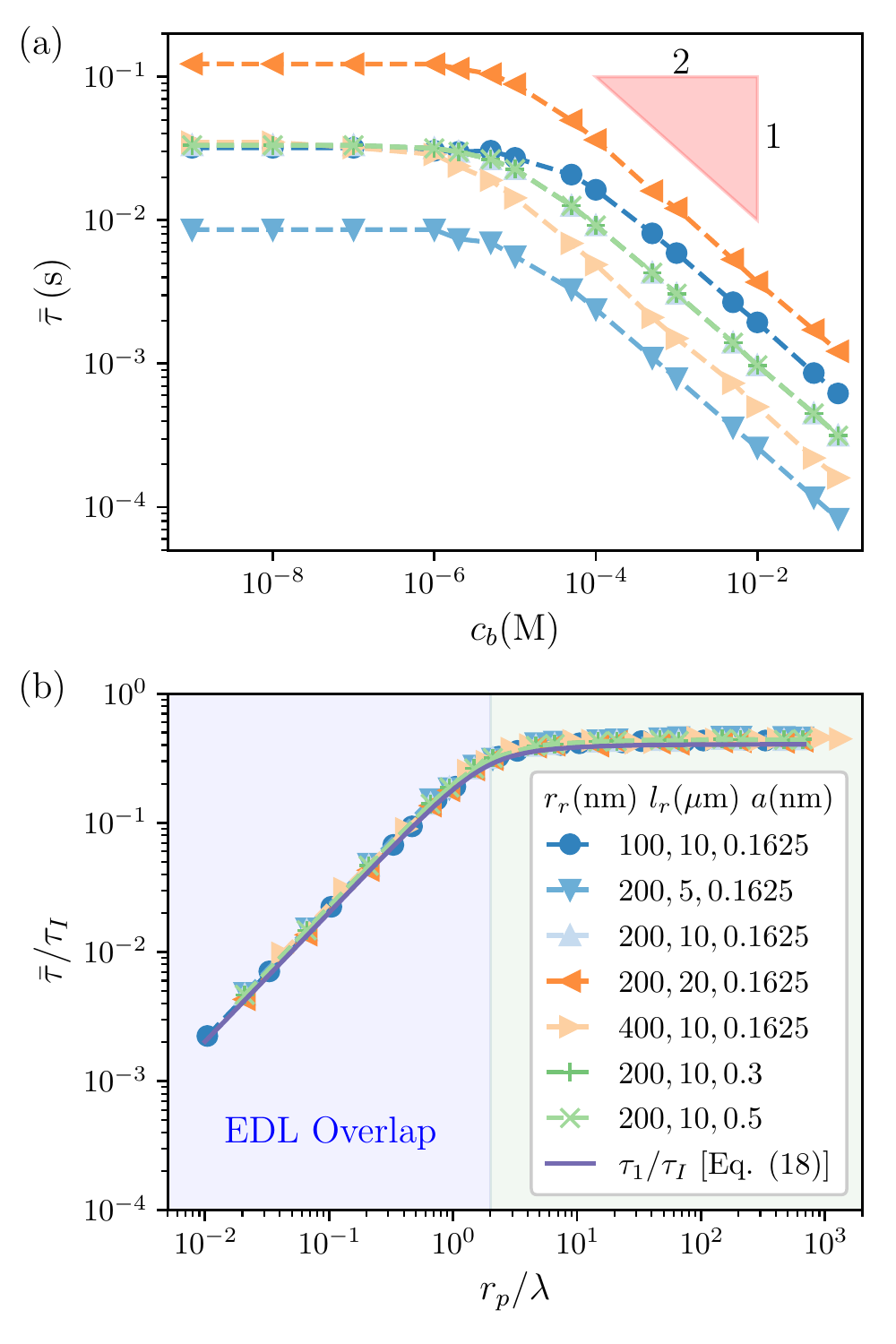}
\caption{
(a) Numerical relaxation $\bar{\tau}$ of our pore setup for $\Phi=10^{-3}$ and various pore sizes and diffusion coefficient $D$, plotted against $c_{b}$. 
Panel (b) shows the same data as (a), normalized by $\tau_{I}$ and plotted against $r_{p}/\ld$. 
The reservoir size was set to $l_{r}=\SI{10}{\micro\meter}$ and $r_{r}=\SI{10}{\micro\meter}$. The legend in panel b also applied to panel a.}
\label{fig:scale} 
\end{figure}

\begin{figure}
\includegraphics[width=8.6cm]{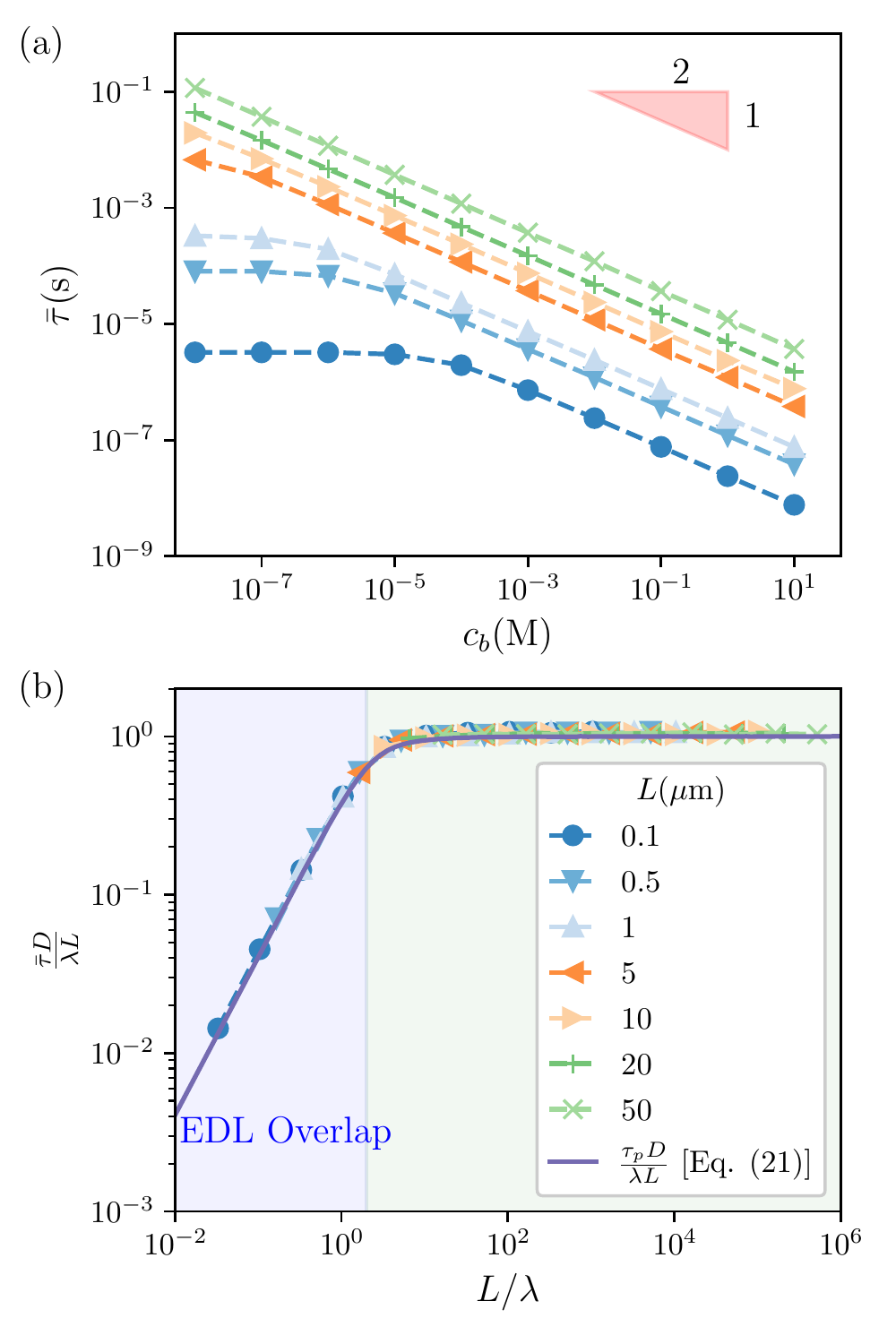}
\caption{
(a) Numerical relaxation $\bar{\tau}$ for planar electrodes subject to $\Phi=10^{-3}$ with various separated distance $L$ plotted against $c_{b}$. 
(b) The same data in (a) normalized by $\ld L/D$, plotted against $L/\ld$.}
\label{fig:plane}
\end{figure}

We then drew figures similar to \cref{fig:c} for cases wherein we varied $\ell_p$, $r_p$, and $a$, see panels (a), (d), and (g) of \cref{fig:geom} of the Supplementary Materials.
From these data, we determined the respective numerical charging timescales $\bar{\tau}$, which we collect in \cref{fig:scale}(a).
We see there that $\bar{\tau}$ is independent of $c_{b}$ for dilute electrolytes, while $\bar{\tau}\sim c_{b}^{-0.5}$ for concentrated electrolytes.
\cref{fig:scale}(b) presents the same $\bar{\tau}$ data, normalized by $\tau_I$ [\cref{eq:deftimescales}] and now versus $r_{p}/\ld$.
With this scaling, data for the different pore sizes and ionic diameters collapse onto a single curve that accurately agrees with $\tau_{1}/\tau_I$ from \cref{eq:tau1RRbinf}.
To understand \cref{fig:scale}(b) qualitatively, note that the ratio of Bessel functions in \cref{eq:tau1RRbinf} behaves as
\begin{align}
\frac{I_{1}\left(r_{p}/\ld\right)}{I_{0}\left(r_{p}/\ld\right)}=
\begin{cases}
\frac{r_p}{2\ld}, \qquad &{\rm for}\quad r_{p}\ll\ld\\
1, \qquad &{\rm for}\quad r_{p}\gg\ld\,.
\end{cases}
\end{align}
With \cref{eq:tau1RRbinf} we then find
\begin{align}\label{eq:t1}
\frac{\tau_{1}}{\tau_{I}}=\frac{4}{\pi^2}\times
\begin{cases}
\frac{r_p}{2\ld}, \qquad &{\rm for}\quad r_{p}\ll\ld\\
1, \qquad &{\rm for}\quad r_{p}\gg\ld\,,
\end{cases}
\end{align}
which agrees with the scaling observed in \cref{fig:scale}(b).

The $r_{p}/\ld$-dependent charging dynamics of our pore-reservoir-pore setup is reminiscent of the charging of an electrolyte between two {\it planar} electrodes separated by a distance $L$---for which $L/\ld$ is a key parameter.
For the latter setup, the linearized PNP equations can be solved with a Laplace transformation, which was first done approximately by Bazant, Thornton, and Ajdari \cite{bazant2004diffuse} and later exactly by Janssen and Bier \cite{janssen_transient_2018} and Palaia \cite{palaia2019charged}.
In particular, \cit{janssen_transient_2018} predicted the following late-time relaxation timescale:
\begin{subequations}\label{eq:janssenbier}
\begin{align}
\tau_p&=\frac{\ld^2}{D(1+\mathcal{M}_1^2\ld^2/L^2)}\,,\\
\mathcal{M}_{1}&\equiv\begin{cases}
M_1, \qquad &{\rm for}\quad L/\ld < \sqrt{3}\,,\\
i\tilde{m}_1, \qquad &{\rm for}\quad L/\ld > \sqrt{3}\,,
\end{cases} 
\intertext{where $M_1$ and $\tilde{m}_1$ are the smallest solutions of two transcendental equations,}
\tan M&=M\left(1+M^2 \ld^2/L^2\right)\,,\\
\tanh \tilde{m}&=\tilde{m}\left(1-\tilde{m}^2 \ld^2/L^2\right)\,.
\end{align}
\end{subequations}
\cref{eq:janssenbier} has the following limiting behavior:
\begin{align}\label{eq:plane}
\tau_{p}=
\begin{cases}
\frac{4L^2}{\pi^2D}[1+O(L/\ld)^2], \qquad &{\rm for}\quad L/\ld \ll\sqrt{3}\,,\\
\frac{\ld L}{D}[1+O(\ld/L)], \qquad &{\rm for}\quad L/\ld \gg \sqrt{3}\,,
\end{cases}
\end{align}
For four values between $L/\ld=11$ and $32$, Asta and coworkers \cite{asta_lattice_2019} showed with Lattice Boltzmann Electrokinetics simulations that \cref{eq:janssenbier} predicted the relaxation timescale more accurately than the well-known $RC$ time $\ld L/D$. 
To our knowledge, the predictions of Refs.~\cite{janssen_transient_2018,palaia2019charged} for $L/\ld<1$ have not been numerically tested.
Therefore, we used the same MPNP implementation as before to simulate the charging dynamics of two flat plates over a wide range of $L$ and $c_{b}$.
\cref{fig:plane}(a) shows numerical results for the numerical charging timescale $\bar{\tau}$.
We observe that $\bar{\tau}\sim c_{b}^{-0.5}$ for most cases except for extremely dilute electrolyte in narrow confinement. 
\cref{fig:plane}(b) shows that the same data collapse onto a single curve when we scale $\bar{\tau}$ by $\ld L/D$ and plot these data against $L/\ld$.
The data (symbols) in this panel agree excellently with the theoretical prediction of \cref{eq:janssenbier} (line).

\begin{figure}
\includegraphics[width=8.6cm]{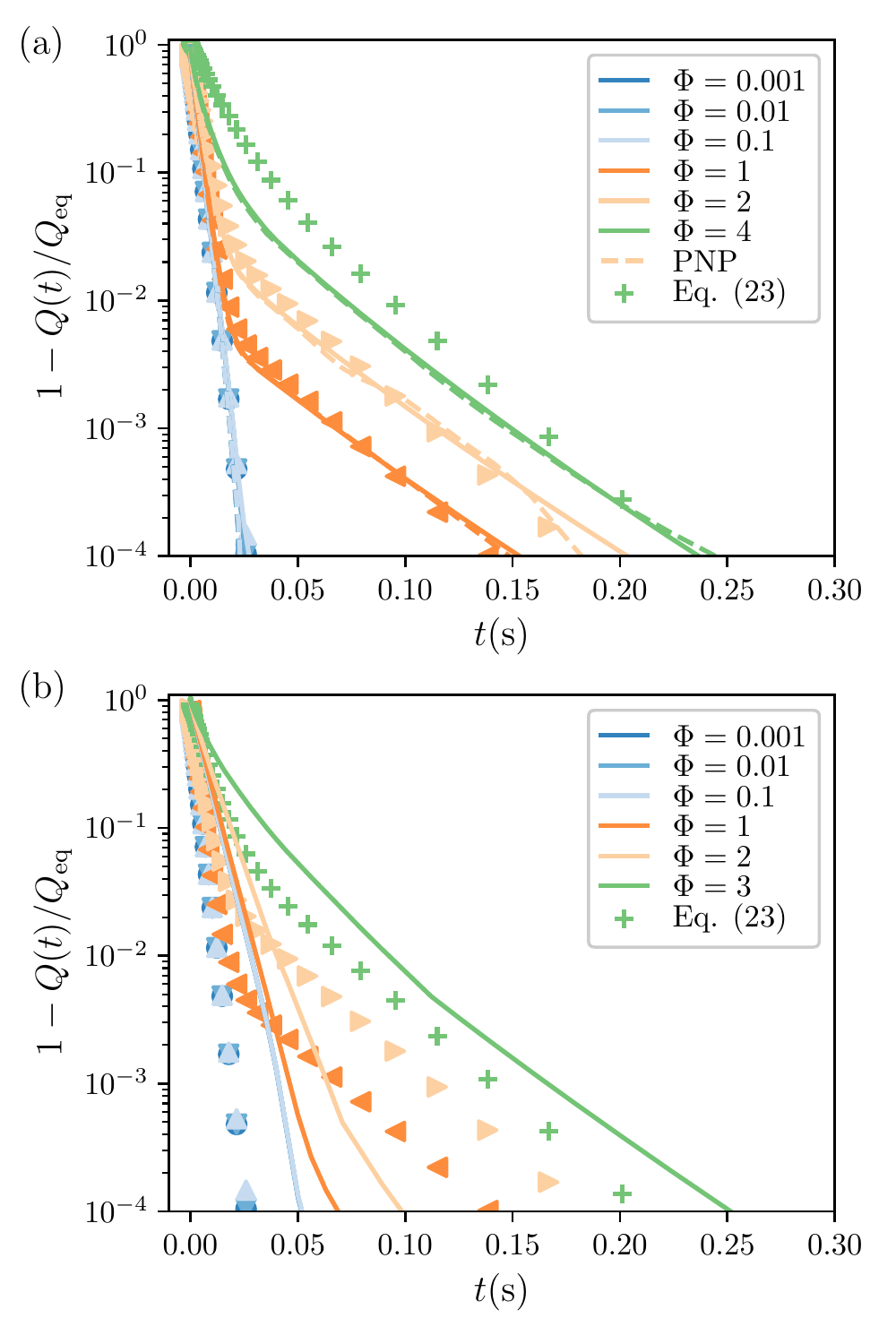}
\caption{
Surface charge relaxation at different values of applied potentials $\Phi$ and for different reservoir sizes of (a) $r_{r}=~\SI{10}{\micro\meter}$ and (b) $r_{r}=~\SI{200}{\nano\meter}$. 
Lines represent results from MPNP, dashed lines represent results from PNP, and dots portray \cref{eq:PNP_charge_dynamics_late_time}.
Parameters are set to $\ell_{p}=\SI{10}{\micro\meter}$, $a=\SI{0.1625}{\nano\meter}$, $r_{p}=\SI{200}{\nano\meter}$, and $c_{b}=~\SI{e-3}{\molar}$, such that $r_p/\ld=20.7$, i.e., nonoverlapping EDLs.} 
\label{fig:vol} 
\end{figure}

\section{Charging at moderate applied potentials $\Phi\sim1$}
Porous electrodes subject to moderate to large potentials are known to acquire charge ``biexponentially", that is, the surface charge is a sum of (at least) two exponential functions with two different timescales \cite{biesheuvel2010nonlinear,kondrat2014NM,janssen2017coulometry, lian2020blessing,breitsprecher2018charge, aslyamov2022relation}. 
From the modeling point of view, relaxation of porous electrodes on two timescales was first predicted by Biesheuvel and Bazant \cite{biesheuvel2010nonlinear}.
Mirzadeh and coworkers performed DNS of the PNP equations and found the effect of biexponential charge buildup---namely, charging slowdown---but did not disentangle the two exponential regimes.
Aslyamov and Janssen \cite{aslyamov2022analytical} studied a slit pore of width $H$ with thin EDLs ($H\gg\ld$), for which they derived
\begin{align}\label{eq:PNP_charge_dynamics_late_time}
\frac{Q(t)}{Q_{\rm eq}}&\simeq1-\frac{8}{\pi^2}\left[\exp{\left(-\frac{\pi^2 }{4}\frac{1}{\cosh\left(\Phi/2\right)}\frac{t}{\tau_I}\right)}\right.\nn
&\qquad\left.+\frac{4\ld}{H}\sinh^2\left(\frac{\Phi}{4}\right)\exp{\left(-\frac{\pi^2}{4}\frac{t}{\tau_{II}}\right)}\right] +O(\eta^2)\,,
\end{align}
where the discarded higher-order terms involve a Dukhin number
\begin{equation}\label{eq:Dukhin}
\eta=4\exp\left(\frac{\Phi}{2}\right)\frac{\lambda}{r_p}\,.
\end{equation}
For the thin EDLs considered in \cit{aslyamov2022analytical}, $\tau_{II}\gg\tau_{I}$ [cf.~\cref{eq:deftimescales}], which means that \cref{eq:PNP_charge_dynamics_late_time} predicts relaxation on two well-separated timescales (unless $\Phi\gg1$).
The second exponential term goes with exactly the same timescale as we found in \cref{eq:henrique,eq:t1}, though its origin is now the moderate applied potential rather overlapping EDLs.
Note that, in \cref{eq:PNP_charge_dynamics_late_time,eq:Dukhin}, we replaced the pore width $H$ of \cit{aslyamov2022analytical} by our pore radius $r_p$.
We did this because a slit and a cylindrical pore have hydraulic radii $H/2$ and $r_p/2$ \cite{MIRZADEH2014633}, respectively, so that $H$ and $r_p$ play similar roles.

\cref{fig:vol} shows the charge buildup of our setup (lines) for $\Phi =0.001,0.01,0.1,1,2$, and $4$ for a wide reservoir ($r_{r}=\SI{10}{\micro\meter}$) (a) and a narrow reservoir ($r_{r}=\SI{200}{\nano\meter}$) (b) as determined with DNS of the MPNP equations.
We also plot \cref{eq:PNP_charge_dynamics_late_time} (symbols) for the same $\Phi$. 
For the wide reservoir [\cref{fig:vol}(a)], the numerics agree with \cref{eq:PNP_charge_dynamics_late_time} well except for $\Phi=4$.
We see that, up to about $t=\SI{0.05}{s}$, $1-Q(t)/Q_{\rm eq}$ relaxes exponentially with a $\Phi$-dependent slope, in agreement with the first line of \cref{eq:PNP_charge_dynamics_late_time}.
(For $\Phi=4$, the slow down is less than predicted.)
Thereafter, a second, slower exponential relaxation emerges which becomes more important with increasing $\Phi$, in line with the $\sinh^2\left(\Phi/4\right)$ term in \cref{eq:PNP_charge_dynamics_late_time}.

\cref{eq:PNP_charge_dynamics_late_time} was derived from the PNP equations, whereas our DNS dealt with MPNP.
For comparison, we also show DNS of the PNP equations [\cref{eq:MPNP} without the last term of \cref{eq:mNP}] with dashed lines in \cref{fig:vol}(a).
The data for $1- Q(t)/Q_{\rm eq}$ is almost the same for PNP and MPNP.
This is not surprising as, for the $a=\SI{0.1625}{\nano\meter}$ and $c_{b}=\SI{e-3}{\molar}$ considered here, we have volume fraction $v=2a^3 c_b=\SI{5.17e-6}{}$; 
Fig.~(5) of \cite{kilic2007stericI} shows that the capacitance of modified and regular Poisson Boltzmann theory hardly differ for $\Phi<10$ for such a small $v$.
Concluding, the difference between PNP and MPNP does not explain the discrepancy between the dots and lines in \cref{fig:vol}(a) at $\Phi=4$.

From \cref{eq:Dukhin} we see that the accuracy of \cref{eq:PNP_charge_dynamics_late_time} depends both on the surface potential and the EDL overlap.
For $\Phi=2$ we find the smallish Dukhin number $\eta=0.52$, which explains the decent agreement between theory and DNS observed in \cref{fig:vol}(a) for that $\Phi$ value.
Conversely, $\Phi=4$ yields $\eta=1.43$, and $O(\eta^2)$ terms are thus no-longer small compared to the other terms in \cref{eq:PNP_charge_dynamics_late_time}, which are of $O(\eta)$ and $O(1)$.
This explains the discrepancies in \cref{fig:vol}(a) between \cref{eq:PNP_charge_dynamics_late_time} and the DNS at $\Phi=4$.
As $\eta\propto c_b^{-1/2}$, one would expect the agreement between \cref{eq:PNP_charge_dynamics_late_time} and the DNS to improve with increasing $c_b$, which we indeed observe below (cf.~\cref{fig:vol2}).

\begin{figure}
\includegraphics[width=8.6cm]{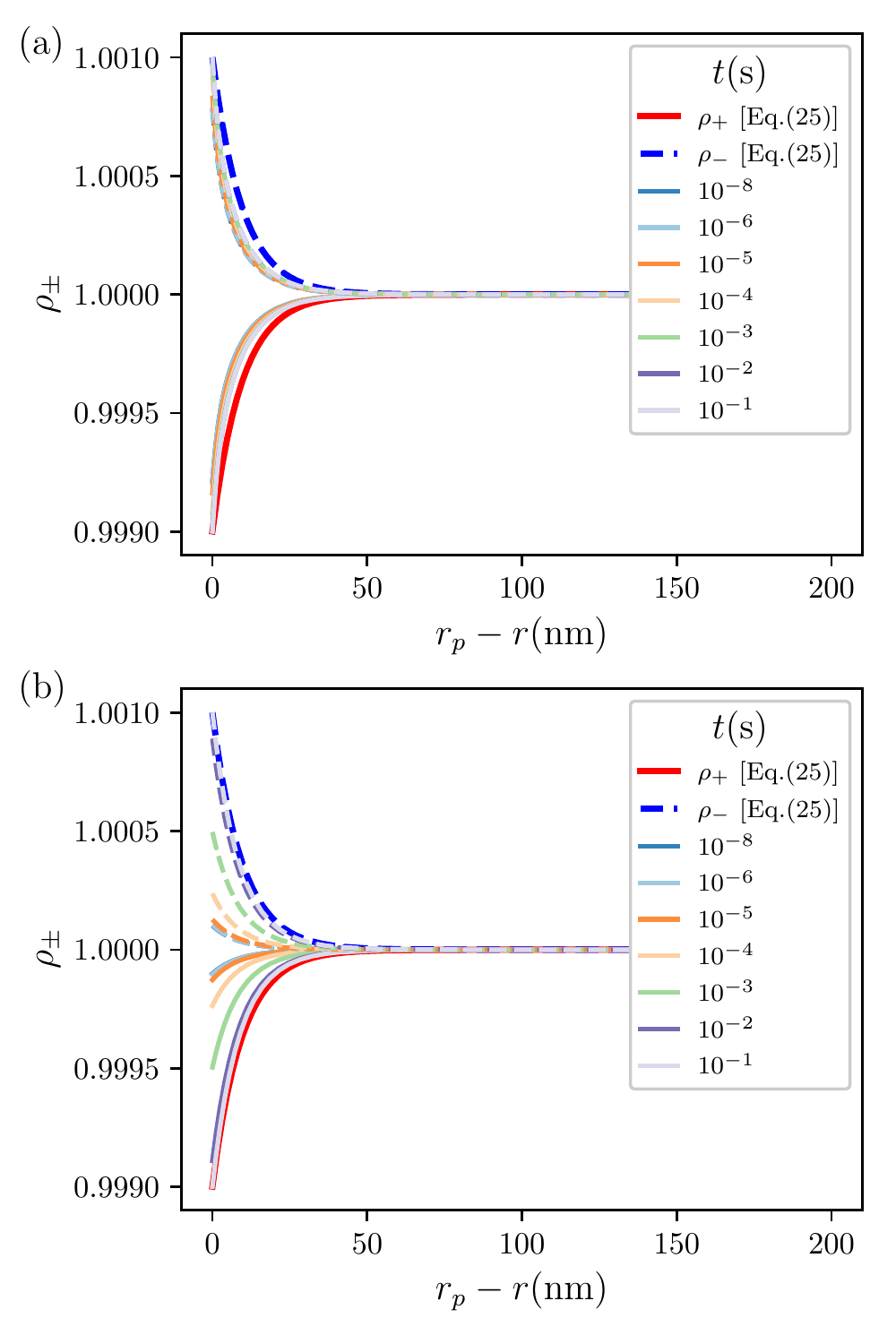}
\caption{
Evolution of normalized cation (solid lines) and anion (dashed lines) densities along the $r$ axis at the orifice for different reservoir size of (a) $r_{r}=~\SI{10}{\micro\meter}$ and (b) $r_{r}=~\SI{200}{\nano\meter}$ under applied potentials $\Phi$ for electrolyte concentration of $c_{b}=~\SI{e-3}{\molar}$.
The other parameters are set to $\ell_{p}=\SI{10}{\micro\meter}$, $r_{p}=\SI{200}{\nano\meter}$, $\ell_p=\SI{10}{\micro\meter}$, $\Phi=10^{-3}$ and $a=\SI{0.1625}{\nano\meter}$.} 
\label{fig:ions} 
\end{figure}

\begin{figure*}
\includegraphics[width=17.8cm]{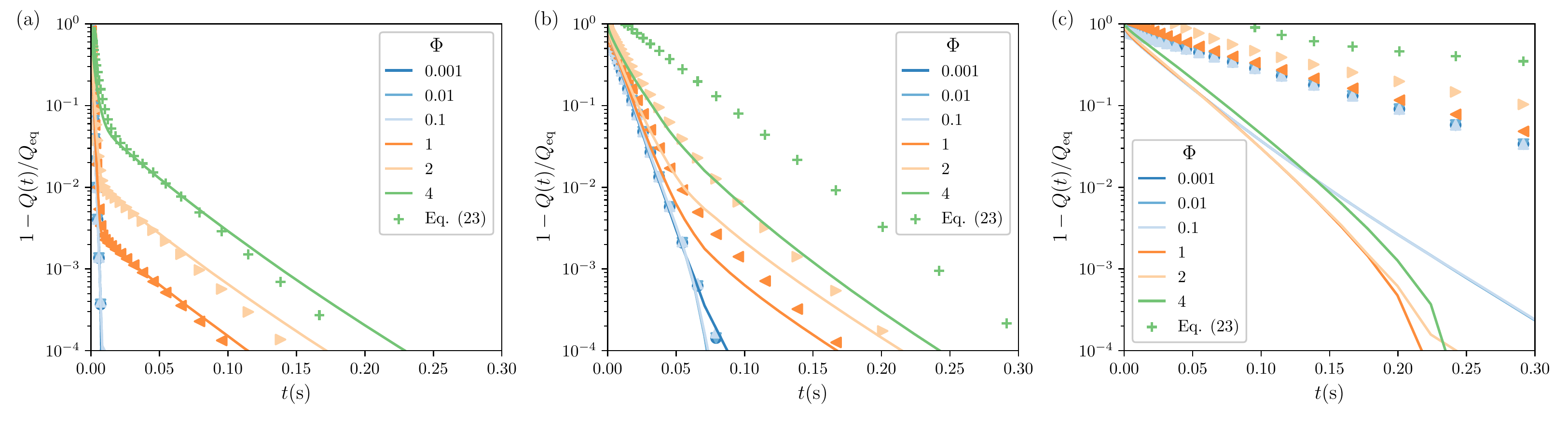}
\caption{
Surface charge relaxation at different values of applied potentials $\Phi$ at electrolyte concentrations of (a) \SI{e-2}{\molar}, (b)~\SI{e-4}{\molar}, and (c) \SI{e-6}{\molar}. 
The other parameters are set to $\ell_{p}=\SI{10}{\micro\meter}$, $r_{p}=\SI{200}{\nano\meter}$, and $a=\SI{0.1625}{\nano\meter}$.}
\label{fig:vol2}
\end{figure*}

For the narrow reservoir ($R/R_b=2$), the agreement in \cref{fig:vol}(b) between the numerics and \cref{eq:PNP_charge_dynamics_late_time} is much worse than in \cref{fig:vol}(a).
This was already anticipated in \cit{aslyamov2022analytical}.
The model therein did not explicitly treat the reservoir but instead postulated the ionic number density at the pore mouth ($z=0$) to instantaneously adapt to the equilibrium Gouy-Chapman solution
\begin{equation}\label{eq:GC_density}
\rho_{\pm}(x)=\left(\frac{1+\tanh(\Phi/2)\exp(-x/\ld)}{1-\tanh(\Phi/2)\exp(-x/\ld)}\right)^{\mp2}\,,
\end{equation}
with $x$ the distance from the electrode surface.
Reference~\cite{aslyamov2022analytical} suggested that this postulate would work better the larger $R/R_b$---which we indeed observe now in \cref{fig:vol}---as this implies that the reservoir is essentially in quasi-equilibrium while the pore charges.
To explicitly check the validity of the postulate in \cit{aslyamov2022analytical}, in \cref{fig:ions}, we compare \cref{eq:GC_density} for $x=r_{p}-r$ to MPNP density profiles at the orifice ($z=0$) for the case of (a) a wide and (b) a narrow reservoir.
We see that $\rho_\pm(r_p-r,z=0,t)$ indeed approach their steady-state profiles much faster for the wide than for the narrow reservoir.
For the wide reservoir, the density profiles at the orifice are almost equilibrated at $t=\SI{e-8}{\second}$, while the rest of the pore relaxes five orders of magnitude slower with $\tau_1= \SI{2.89e-3}{\second}$.
From the point of view of the rest of the pore, the orifice thus relaxes instantaneously.
Last, we note that the late-time ion densities are closer to the Gouy-Chapman prediction for the narrow than for the wide reservoir. 
While postulating instantaneously-relaxed ion densities at the orifice may thus be justified when $R\gg R_b$, these densities may deviate slightly from those deeper in the pore.

In \cref{fig:vol2} we again consider various potentials, now for three different $c_b$.
As anticipated, \cref{eq:PNP_charge_dynamics_late_time} describes the DNS better at higher $c_b$.
For $c_{b}=10^{-2}~\si{M}$, we see in \cref{fig:vol2}(a) that the pore relaxes biexponentially with two vastly different timescales. 
Here, \cref{eq:PNP_charge_dynamics_late_time} describes the DNS even at $\Phi=4$, for which, now, $\eta=0.45$ is indeed still smallish.
For $c_{b}=10^{-4}~\si{M}$, we see in \cref{fig:vol2}(b) that the pore still relaxes biexponentially, but that two timescales differ less than for $c_{b}=10^{-6}~\si{M}$ [\cref{fig:vol2}(c)]. 
We understand this with \cref{eq:PNP_charge_dynamics_late_time}, wherein $\tau_I$ decreases with $c_b$, while $\tau_{II}$ does not depend on it.
For $c_{b}=10^{-6}~\si{M}$, we see predictions from DNS and from \cref{eq:PNP_charge_dynamics_late_time} for $1-Q(t)/Q_{\rm eq}$ do not agree at all.

\section{Conclusions}
Through direct numerical simulation (DNS) of the modified Poisson-Nernst-Planck (MPNP) equations, we have studied the charging dynamics of two cylindrical electrolyte-filled pores on either side of a cylindrical electrolyte reservoir, subject to a sudden potential difference.
The pores charge exponentially with different timescales, whose dependence on the various system parameters we scrutinized.

For small applied potentials, we found quantitative agreement between our DNS of the MPNP equations and the analytical result by Janssen \cite{janssen2021transmission} for the bulk-resistance dependence of the TL timescale, both for overlapping and nonoverlapping EDLs.
We showed that, contrary to conventional wisdom \cite{biesheuvel2010nonlinear,henrique2021charging}, the potential in the reservoir is not linear when the reservoir is wider than the pore: it decays much faster into the reservoir.
We also discussed the influence of the reservoir resistance on the early-time charging behavior of our system: for $R/R_b\gg1$, we recovered the known $Q\propto \sqrt{t}$ charging of \cit{sakaguchi2007}; for $R/R_b\sim1$, we found a new linear scaling behavior $Q\propto t$.
In several ways, our work thus highlights the importance of the electrolyte reservoir on the pore's charging dynamics, which was ignored in many prior studies.
Further, we compared Posey and Morozumi's TL equation solution to DNS of the MPNP equations and found that their solution generally works well at late times and in the interior of the pore; differences between the DNS and TL model were visible at early times and especially near the pore's orifice.
Future TL models should thus pay close attention to the boundary and initial conditions used.

For moderately strong applied potentials, we compared our DNS to a recent theoretical prediction of Aslyamov and Janssen \cite{aslyamov2022analytical}.
We found good agreement between these methods for small Dukhin numbers $\eta$, but only if the pore resistance $R$ was vastly greater than the reservoir resistance $R_b$.
Discrepancies between these methods for $R\sim R_b$ were traced to the postulate in \cit{aslyamov2022analytical} that the density profiles at the pore's orifice relax instantaneously, which we showed to be reasonable only for $R\gg R_b$.
Future work could thus try to generalize the findings of \cit{aslyamov2022analytical} for cases where $R\sim R_b$.

We hope that the insights from our numerical study motivate further work, not only on improved theoretical models, but also on new experiments that probe porous electrode charging at the single-pore level.

J.Y. and M.J. contributed equally to this work. 
This work is part of the D-ITP consortium, a program of the Netherlands Organization for Scientific Research (NWO) that is funded by the Dutch Ministry of Education, Culture and Science (OCW). 
We acknowledge the EU-FET project NANOPHLOW (REP-766972-1) and helpful discussion with Prof. Honglai Liu and Willem Boon.
We thank Timur Aslyamov for his useful comments on our manuscript.

\bibliographystyle{apsrev4-2}
\bibliography{main.bib}

\clearpage
\setcounter{figure}{0}
\setcounter{section}{0} 
\renewcommand{\thefigure}{{S}\arabic{figure}}
\renewcommand{\thesection}{{S}\arabic{section}}
\newpage
\onecolumngrid
\begin{center}
{\bf SUPPLEMENTARY MATERIAL to: Direct numerical simulations of the modified Poisson-Nernst-Planck equations for the charging dynamics of cylindrical electrolyte-filled pores}\\\vspace{0.2cm}
Jie Yang, Mathijs Janssen, Cheng Lian, and Ren\'{e} van Roij
\end{center}

\date{\today}

\twocolumngrid

\section{Influence of the cap with rounded edged}\label{appendix:cap}
\begin{figure}[b]
\includegraphics[width=8.6cm]{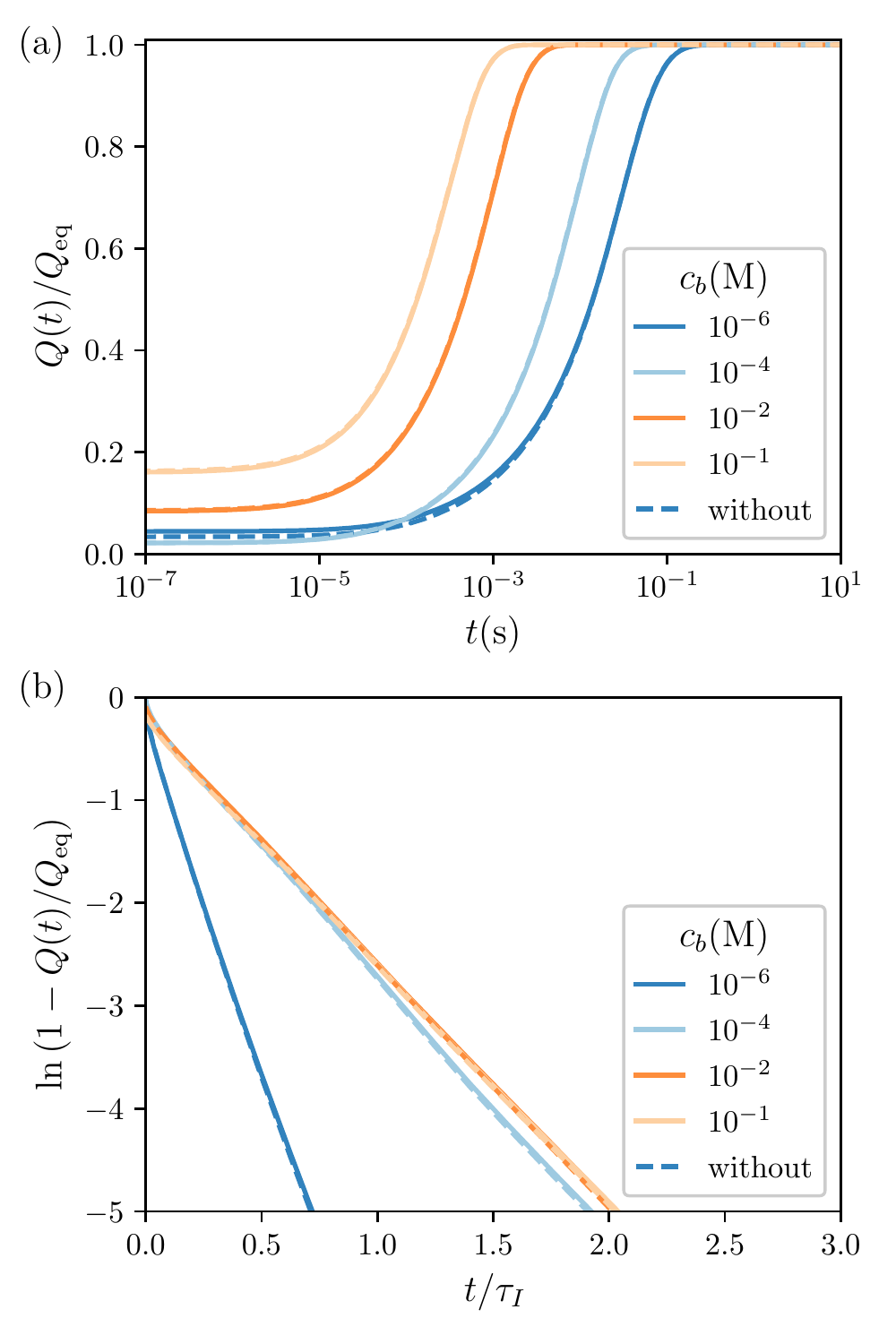}
\caption{
(a) The surface charge density $Q(t)/Q_{\rm eq}$ versus time, (b) surface charge relaxation versus time scaled by $\tau_{\rm int}$ for various electrolyte concentrations with (solid lines) or without (dashed lines) adding the cap with rounded edges, with $\tau_{I}$ the TL time given by \cref{eq:deftimescales}. 
For both panels, $\ell_{p}=\SI{10}{\micro\meter}$, $r_{p}=\SI{200}{\nano\meter}$, $a = \SI{0.1625}{\nano\meter}$, $\Phi=10^{-3}$, $l_{r}=\SI{10}{\micro\meter}$, and $r_{r}=\SI{10}{\micro\meter}$.}
\label{fig:semi}
\end{figure}

\cref{fig:semi} compares the charging with or without the connection regions and the caps at the end of the pores. 
We see that adding these regions has no substantial influence on the charging.

\section{Parametric dependence of pore charging}
We discuss the dependence of charging on various system parameters. 
The data presented here was used to draw \cref{fig:scale} of the main text.

\begin{figure*}
\includegraphics[width=17.8cm]{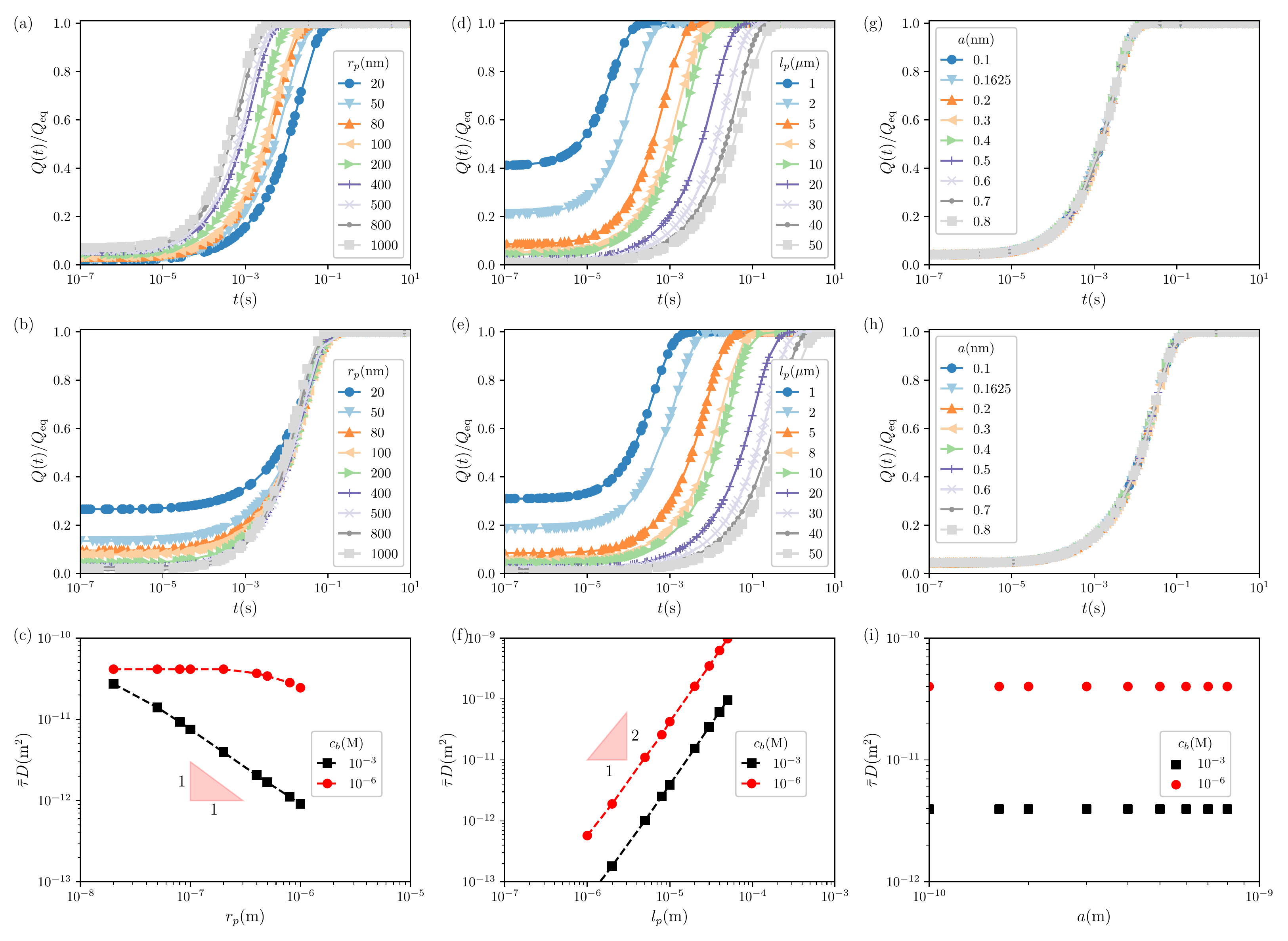}
\caption{
Effects of (a-c) pore radius $r_{p}$, (d-f) pore length $\ell_{p}$, and (g-i) ionic diameter $a$ on pore charging dynamics. 
For different values of these parameters, we show the normalized surface charge density $Q(t)/Q_{\rm eq}$ versus time for (a,d,g) $c_{b}=\SI{e-3}{\molar}$ and (b,e,h) $c_{b}=\SI{e-6}{\molar}$.
From these data, we determined the numerical timescale $\bar{\tau}$ as a function of (c) $r_{p}$, (f) $\ell_{p}$, and (i) $a$ for $c_{b}=\SI{e-3}{\molar}$ (black squares) and $c_{b}=\SI{e-6}{\molar}$ (red circles).
In all panels, $\Phi=10^{-3}$ and $l_{r}=r_{r}=\SI{10}{\micro\meter}$.
We further used (a-c, g-i) $\ell_{p}=\SI{10}{\micro\meter}$, (d-i) $r_{p}=\SI{200}{\nano\meter}$, and (a-f) $a=\SI{0.1625}{\nano\meter}$.}
\label{fig:geom}
\end{figure*}

First, we study pore charging for different pore radii $r_{p}$. 
\cref{fig:geom}(a) and (b) show the time-dependent scaled surface charge $Q(t)/Q_{\rm eq}$ for (a) $c_{b}=\SI{e-3}{\molar}$ and (b) $c_{b}=\SI{e-6}{\molar}$.
For $c_{b}=\SI{e-3}{\molar}$, EDLs in the pore are nonoverlapping for all $r_p$. 
The data in \cref{fig:geom}(a) for this case show that charging goes faster for wider pores. 
For $c_{b}=\SI{e-6}{\molar}$, EDLs are overlapping for the smaller $r_p$ considered. 
The data in \cref{fig:geom}(b) for this case collapse below $r_p=200$.
Next, we plot the numerical timescales calculated from the same data above as a function of $r_{p}$ [\cref{fig:geom}(c)]. 
In agreement with the two limiting regimes of \cref{eq:t1}, the numerical charging timescale $\bar{\tau}$ scales as $1/r_{p}$ for $c_{b}=\SI{e-3}{\molar}$, while, for $c_{b}=\SI{e-6}{\molar}$ it hardly depends $r_{p}$. 

Second, we study pore charging for different pore lengths $\ell_{p}$ and the same $c_{b}=\SI{e-3}{\molar}$ and $c_{b}=\SI{e-6}{\molar}$ as before. 
\cref{fig:geom}(d) and (e) present the normalized surface charge versus time.
These panels show that charging goes slower with increasing $\ell_{p}$. 
The numerical corresponding timescales $\bar{\tau}$ versus the pore length $\ell_{p}$ are presented in the log-log plot \cref{fig:geom}(g).
For both salt concentrations, the slope of the data in \cref{fig:geom}(g) is roughly 2, indicating that $\bar{\tau}\propto \ell_{p}^{2}$, in agreement with both limiting regimes of $\tau_{1}$ in \cref{eq:t1}.

Last, \cref{fig:geom}(h) and (i) present the normalized surface charge variation versus time for different ionic diameters $a$.
These panels show that the charging process is not affected by $a$. 
This is easy to understand: for the small potential $\Phi=10^{-3}$ considered here, MPNP and PNP are essentially the same, and PNP does not depend on $a$.

\end{document}